\documentclass[10pt,twocolumn,twoside]{IEEEtran}
\usepackage{graphicx}
\usepackage{latexsym}
\usepackage{amsmath}
\usepackage{amssymb}
\usepackage{epsfig}
\usepackage{cite}
\usepackage{algorithmic}
\usepackage{overpic}
\usepackage{multirow}
\usepackage[svgnames]{xcolor}
\usepackage{empheq}
\usepackage{colortbl,array}
\usepackage{multirow,bigdelim}
\usepackage{subfigure}
\usepackage{booktabs,amsmath}
\usepackage{fancyhdr, graphicx, amsmath, amssymb}
\usepackage[linesnumbered,ruled]{algorithm2e}
\usepackage{arydshln}
\usepackage[english]{babel}
\usepackage{xcolor}
\usepackage{tcolorbox}
\usepackage{soul}
\usepackage[hidelinks]{hyperref}
\hypersetup{breaklinks=true}
\newtheorem{lem}{Lemma}
\newtheorem{thm}{Theorem}
\newtheorem{prom}{Problem}
\newtheorem{cor}{Corollary}
\newtheorem{asm}{Assumption}
\newtheorem{definition}{Definition}
\newtheorem{rem}{Remark}
\newtheorem{prop}{Proposition}

\colorlet{shadecolor}{FloralWhite!60}
\colorlet{titleshade}{OliveDrab!15}
\newsavebox{\mysaveboxM}
\newsavebox{\mysaveboxT}

\usepackage[linesnumbered,ruled]{algorithm2e}
\usepackage{arydshln}
\usepackage{cite}

\usepackage{mathtools}
\newcommand\iidsim{\stackrel{\mathrm{i.i.d.}}{\sim}}

\usepackage[symbol*]{footmisc}
\DefineFNsymbolsTM{myfnsymbols}{
\textparagraph \mathparagraph
}
\setfnsymbol{myfnsymbols}

\usepackage{bigfoot}

\DeclareNewFootnote[para]{math}[fnsymbol]
\MakeSortedPerPage{footnotemath}

\ifCLASSINFOpdf
\else
\fi

\begin{document}

\title{Social System Inference from Noisy Observations}

\author{Yanbing~Mao, Naira~Hovakimyan, Tarek~Abdelzaher, and Evangelos~Theodorou
\thanks{Y.~Mao and N.~Hovakimyan are with the Department of Mechanical Science and Engineering, University of Illinois at Urbana--Champaign, Urbana, IL 61801, USA (e-mail: \{ybmao, nhovakim\}@illinois.edu).}
\thanks{T.~Abdelzaher is with the Department of Computer Science, University of Illinois at Urbana-Champaign, Urbana, IL 61801, USA (e-mail: zaher@illinois.edu).}
\thanks{E.~Theodorou is with the Department of Aerospace Engineering, Georgia Institute of Technology, Atlanta, GA 30332, USA (e-mail: evangelos.theodorou@gatech.edu).}
\thanks{This work was supported in part by DOD HQ00342110002, DARPA W911NF-17-C-0099, AFOSR FA9550-15-1-0518, and NSF CNS-1932529.}}

\maketitle
\begin{abstract}
This paper studies social system inference from a single trajectory of public evolving opinions, wherein observation noise leads to the statistical dependence of samples on time and coordinates. We first propose a cyber-social system that comprises individuals in a social network and a set of information sources in a cyber layer, whose opinion dynamics explicitly takes confirmation bias, novelty bias and process noise into account. Based on the proposed social model, we then study the sample complexity of least-square auto-regressive model estimation, which governs the number of observations that are sufficient for the identified model to achieve the prescribed levels of accuracy and confidence. Building on the identified social model, we then investigate social inference, with particular focus on the weighted network topology, the subconscious bias and the model parameters of confirmation bias and novelty bias. Finally, the theoretical results and the effectiveness of the proposed social model and inference algorithm are validated by the US Senate Member Ideology data.
\end{abstract}


\IEEEpeerreviewmaketitle

\section{Introduction}\IEEEPARstart{D}{ynamical} network identification from observed nodal states, with a particular focus on graph topology identification/reconstruction,  has gained widespread attention in a wide variety of fields, ranging from power networks \cite{weng2016distributed} to social networks \cite{wai2015social}. In social networks, individuals' bias -- cognitive bias and subconscious bias -- pose a formidable challenge to the model-based network topology identification, since they are as critical as network topology in the public opinion evolution \cite{nickerson1998confirmation,lamberson2018model}. Therefore, a fairly accurate social model of opinion evolution which explicitly takes cognitive bias as well as subconscious bias into account is indispensable for the topology identification with prescribed levels of accuracy and confidence (PAC). Leveraging the opinion evolution model, the network topology,  the cognitive bias and the subconscious bias can be decoupled from each other.


The social model of opinion evolution has been an active subject for decades, among which a few well-known models have been proposed to capture individual conformity, cognitive and subconscious behaviors \cite{proskurnikov2017tutorial,proskurnikov2018tutorial}. For example, DeGroot model \cite{degroot1974reaching} considers individual opinion evolution as an average of her neighbors, which describes conformity behavior. Friedkin-Johnsen model~\cite{friedkin1990social} incorporates individual subconscious bias into opinion evolution, thereby making the model more suitable to several real-life scenarios and  applications \cite{das2013debiasing}. Though imposing a bounded confidence on social influence, Hegselmann-Krause model \cite{hegselmann2002opinion} has the capability of capturing confirmation bias \cite{del2017modeling}, which helps create ``echo chambers" within networks, in which misinformation and polarization thrive \cite{lazer2018science}. Hegselmann-Krause model involves a discontinuity in the influence impact, i.e., an individual completely ignores the opinions that are ``too far" from hers, which renders the steady-state analysis difficult. As a remedy, we proposed an opinion evolution model in \cite{mao2019impact, 8510828}, which is a variation of Friedkin-Johnsen model with continuous and symmetric confirmation bias model. Recently, Abdelzaher et al. in \cite{abdelzaher2020paradox} and Xu et al. in \cite{xu2020paradox} reveal the significant influence of consumer preferences for outlying content on opinion polarization in the modern era of information overload. Meanwhile, Bailey in \cite{lamberson2018model} suggests the novelty bias has the power to override rationalization. Motivated by these discoveries, we incorporate novelty bias -- an influence that tends to distract individuals and turn their attention to shiny, new thing \cite{lamberson2018model} -- to our previously proposed opinion dynamics \cite{mao2019impact, 8510828}. The opinion evolution model in this paper also includes random process noise that describes the model errors and uncertainty.

Ignoring novelty bias, our study of competitive information spread in social networks uncovers the dependence of Nash equilibrium on network topology, subconscious bias and confirmation bias \cite{mao2019impact}; similar discoveries appear in \cite{dhamal2018optimal}.  The studies therein indicate that inferring network topology only is not sufficient for optimal decision making in social networks \cite{wai2015social}. Motivated by this observation, concurrent inference of network topology and confirmation bias are primarily investigated in \cite{9187724}, which however relies on several rather restrictive assumptions: 1) an individual's subconscious bias equates to her initial opinion, 2) observations of evolving opinions are completely reliable, i.e., noise-free, and 3) novelty bias and process noise have no influence on opinion evolution. To remove these assumptions in this paper, we propose a social-system inference procedure, which is based on least-square auto-regressive model estimation. The inference objectives include weighted network topology, public subconscious bias, model parameters of confirmation bias and novelty bias.

Due to process and observation noise, one intuitive question pertaining to the accuracy of social system inference arises: how many observations are sufficient for the inference solution to achieve PAC? To answer this question in the context of system matrix estimation, significant effort has been devoted towards the sample complexity of ordinary least-square estimator in recent \textcolor[rgb]{0.00,0.00,1.00}{few} years \cite{jedra2020finite,simchowitz2018learning,sarkar2019near,sarkar2019finite,8814438}. We note that the analysis of sample complexity therein relies on Hanson-Wright inequality \cite{rudelson2013hanson}, which requires zero-mean, unit-variance, sub-gaussian independent coordinates for noise vectors. Banerjee et al. in \cite{banerjee2019random} considered the generalization of existing results by allowing for statistical dependence on stochastic processes via Johnson--Lindenstrauss transform, which however still requires the noise variables to have zero mean and the marginal random variables to be conditionally independent. In this paper, we reveal that even if the observation noise vectors have $\mathrm{i.i.d}$ coordinates and time, the noise leads to inevitable statistical dependence of opinion observations on time and coordinates. \textcolor[rgb]{0.00,0.00,1.00}{Without additional requirement of $\mathrm{i.i.d}$ and isotropic controlled (random) inputs}, the statistical dependence and non-zero mean of noise hinder the application of obtained sample complexity in \cite{jedra2020finite,simchowitz2018learning,sarkar2019near,sarkar2019finite,8814438} to the social system inference. These observations motivate us to investigate the sample complexity of social system estimation in the presence of observation noise with \textcolor[rgb]{0.00,0.00,1.00}{non-zero} mean, which paves the way for the derivation of social system inference.

The contributions of this paper are summarized as follows.
\begin{itemize}
  \item Building on Friedkin-Johnsen model \cite{friedkin1990social}, we propose an opinion evolution model with incorporation of confirmation bias, novelty bias and process noise.
  \item In the presence of observation noise with non-zero mean, which results in statistical dependence of opinion observations on time and coordinates, we investigate the sample complexity of proposed social system estimation.
  \item Building on social system estimation, we drive a social-system inference procedure for weighted network topology, subconscious bias, model parameters of confirmation bias and novelty bias.
  \item We validate the theoretical results and the effectiveness of proposed opinion evolution model by US Senate Member Ideology data.
\end{itemize}

This paper is organized as follows. In Section II, we present preliminaries. In Sections III and IV, we investigate social system estimation and social system inference, respectively. We present validation results in Section V. We finally present  conclusions in Section VI.

\section{Preliminaries}
\subsection{Notation}
\label{notation}
We let $\mathbb{R}^{n}$ and $\mathbb{R}^{m \times n}$ denote the set of $\emph{n}$-dimensional real vectors and the set of $m \times n$-dimensional real matrices, respectively. $\mathbb{N}$ stands for the set of natural numbers, and $\mathbb{N}_{0} = \mathbb{N} \bigcup {0}$. We let $\mathbf{1}$ and $\mathbf{0}$ denote the vectors of all ones and all zeros, compatible dimensions. We define $\mathbf{I}_{n}$ as $n$-dimension identity matrix. The superscript `$\top$' stands for the matrix transposition. For a matrix $W$, ${\left[ W \right]_{i,j}}$ denotes the element in row $i$ and column $j$. For vectors $x$ and $y$, we let $[x;~y] = [x^{\top}, ~y^{\top}]^{\top}$. \textcolor[rgb]{0.00,0.00,1.00}{The $\sigma$-algebra is denoted by $\sigma(\cdot)$}.

Other important notations are highlighted as follows:
\begin{description}
  \item[$|| A ||:$] ~~~~~spectral norm of matrix $A$;
  \item[$|| A ||_{\mathrm{F}}:$] ~~~~~Frobenius norm of matrix $A$;
  \item[$|| x ||:$] ~~~~~Euclidean norm of vector $x$;
  \item[$\mathbf{E}$] ~~~~~expectation operator;
  \item[$\mathcal{S}^{n-1}:$] ~~~~~unit sphere in $\mathbb{R}^{n}$;
  \item[$\Omega^{\mathrm{c}}:$] ~~~~~complement of event $\Omega$;
  \item[$\mathbf{P}(\Omega):$] ~~~~~probability of event $\Omega$;
  \item[$\lambda _{\min}(A):$] ~~~~~minimum eigenvalue of symmetric matrix $A$.
\end{description}

The social system is composed of $n$ individuals in social network and $m$ information sources. The interaction among individuals is modeled by a digraph $\mathfrak{G} = (\mathbb{V}, \mathbb{E})$, where $\mathbb{V}$ = $\left\{\mathrm{v}_{1}, \ldots,  \mathrm{v}_{n}\right\}$ is the set of vertices representing the individuals, and $\mathbb{E} \subseteq  \mathbb{V} \times \mathbb{V}$ is the set of edges of the digraph $\mathfrak{G}$  representing the influence structure. The communication from information sources to individuals is modeled by a  bipartite digraph $\mathfrak{B} = (\mathbb{V} \bigcup \mathbb{K}, \mathbb{B})$, where $\mathbb{K}$ = $\left\{\mathrm{u}_{1}, \ldots,  \mathrm{u}_{m}\right\}$ is the set of vertices representing information sources, and $\mathbb{B} \subset \mathbb{V} \times \mathbb{K}$ is the set of edges of the digraph.

\vspace{-0.10cm}
\subsection{Social Network Model}
We consider the following model which is adopted from Friedkin-Johnsen model~\cite{friedkin1990social} and its recent variation~\cite{mao2019impact}:
\begin{subequations}
\begin{align}
{x_i}(k \!+\! 1) &\!=\! {\alpha _i}(k){s_i} \!+\! \sum\limits_{j \in \mathbb{V}}\!{{w_{ij}}{x_j}(k)}  \!+\! \sum\limits_{d \in \mathbb{K}} \!{{c}( {{x_i}(k),h_{d}(k)})h_{d}(k)} \nonumber\\
&\hspace{1.5cm} \!+\! \sum\limits_{d \in \mathbb{K}}\!{{n}( {{\bar{x}_i}(k),h_{d}(k)})h_{d}(k)} \!+\!  {\mathfrak{p}_i}(k),\\
{y_i}(k) &\!=\! {x_i}(k) \!+\!{\mathfrak{o}_i}(k) \in [-1,1], ~~i \!\in\! \mathbb{V}, ~k \!\in\! \mathbb{N}.
\end{align}\label{kka}
\end{subequations}
Here we clarify the notations and variables:
\begin{itemize}
  \item ${x_i}\!\left( {k} \right) \in [-1,1]$ is individual $\mathrm{v}_{i}$'s opinion, ${y_i}( {k}) \in [-1,1]$ is \textcolor[rgb]{0.00,0.00,1.00}{observed her} opinion for inference; ${s_i} \in [-1,1]$ is her subconscious bias, which is based on inherent personal characteristics (e.g., socio-economic
      conditions where the individual grew up and/or lives in); ${h_d}(k) \in [-1,1]$ is information source $\mathrm{u}_{d}$'s opinion at time $k$.
  \item ${\mathfrak{p}}_{i}(k)$ denotes process noise  due to model error and uncertainty, ${\mathfrak{o}}_{i}(k)$ denotes observation noise.
  \item $w_{ij}$ represents the influence of individual $\mathrm{v}_{j}$ on  $\mathrm{v}_{i}$, and
  \begin{equation}
	w_{ij} = \begin{cases}
		> 0, & \text{if } (\mathrm{v}_i,\mathrm{v}_j) \in \mathbb{E}\\
		= 0, & \text{otherwise}.
	\end{cases} \nonumber
  \end{equation}
 We note the individual-individual influence weights $w_{ij}$s are cognition- or knowledge-trust based and thus fixed over time, since the cognitive factors that can influence trust decisions are founded on a deeper knowledge of the other person and the stability of the other's behavior across time and contexts, \textcolor[rgb]{0.00,0.00,1.00}{which
tends to vary little over a long period of time \cite{borum2010science}}.
  \item The state-dependent influence weight $\textcolor[rgb]{0.00,0.00,1.00}{c( {{x_i}(k),h_{d}(k)})}$ models symmetric ``confirmation bias" as
      \begin{align}
  \!\!\textcolor[rgb]{0.00,0.00,1.00}{c( {{x_i}(k),h_{d}(k)})} \!=\! 2\epsilon_{i} \!-\! \epsilon_{i} | x_i(k) \!-\! h_{d}(k) |, ~\epsilon_{i} \!\geq\! 0. \label{eq: sdw1}
  \end{align}
  We note that function \eqref{eq: sdw1} can also model homophily \cite{mas2014cultural, duggins2014psychologically}. It is used in this paper to describe the symmetric confirmation bias, whose motivations are: 1) both polarization and homogeneity are the results of the conjugate effect of confirmation bias and social influence \cite{del2017modeling,del2016spreading}, 2) confirmation bias happens when a person gives more weight to evidence that confirms their beliefs and undervalues evidence that could disprove it \cite{CCBB}.
  \item The state-dependent influence weight $\textcolor[rgb]{0.00,0.00,1.00}{n( {{\bar{x}_i}(k),h_{d}(k)})}$ models ``novelty bias" as
      \begin{align}
  \textcolor[rgb]{0.00,0.00,1.00}{n( {{\bar{x}_i}(k),h_{d}(k)})}= \eta_{i} | \bar{x}_i(k) - h_{d}(k) |, ~\eta_{i} \geq 0, \label{sdw1}
  \end{align}
  where ${{\bar x}_i}( k )$ denotes individual $\mathrm{v}_{i}$'s sensed expectation from her neighbors, defined as the mean of her neighbors' opinions, i.e.
\begin{align}
{{\bar x}_i}( k ) \triangleq \frac{1}{{\sum\limits_{l \in \mathbb{V}} { {{w_{il}}}} }}\sum\limits_{j \in \mathbb{V}} {{{w_{ij}}}{x_j}( k )}. \label{sdw2}
\end{align}
 The motivation behind the models \eqref{sdw1} and \eqref{sdw2} can be explained by the significant impact of consumer preferences for outlying content on opinion polarization in the era of information overload revealed in \cite{abdelzaher2020paradox, xu2020paradox}, the well-known fact that novel/outlying information is far away from expectations (which motivates the opinion distance between sensed expectation and opinion of information source in \eqref{sdw1}), and the novelty bias refers to an influence that tends to distract individuals and turn their attention to shiny, new thing \cite{lamberson2018model} (which motivates the strictly increasing function \eqref{sdw1} w.r.t. opinion distance, if $\eta_{i} > 0$).
\item  $\alpha_i(k) \geq 0$  is the ``resistance parameter'' of individual $\mathrm{v}_{\mathrm{i}}$. To guarantee $x_{i}(k) \in [-1, 1]$ for $\forall k \in \mathbb{N}$ and $\forall i \in \mathbb{V}$, it is determined in such a way that
\begin{align}
&\textcolor[rgb]{0.00,0.00,1.00}{{\alpha _i}(k)} + \sum\limits_{j \in \mathbb{V}} {{w_{ij}}}  + \sum\limits_{d \in \mathbb{I}} {{c}( {{x_i}(k),h_{d}(k)})} + \chi_{i}  \nonumber\\
&\hspace{2.31cm}+ \sum\limits_{d \in \mathbb{I}} {{n}( {{\bar{x}_i}(k),h_{d}(k)})}  = 1, \forall i \!\in\! \mathbb{V} \label{sdw3}
\end{align}
where $\chi_{i}$ denotes the bound on process noise in this paper, i.e. $|{\mathfrak{p}_i}(k)| < \chi_{i} \leq 1$, $\forall k \in \mathbb{N}$.
\item Another motivation behind the state-dependent influence (i.e., $c( {{x_i}(k),h_{d}(k)})$ in \eqref{eq: sdw1} and $n( {{\bar{x}_i}(k),h_{d}(k)})$ in \eqref{sdw1}) of information sources on individuals is that information sources lack the rational basis of trust, e.g., news media prioritizes outlying information for attentiveness.
\end{itemize}

\subsection{Inference Objectives}
We denote the inference solution by
\begin{align}
\mathcal{S} \triangleq (\breve{W},~\breve{s},~\breve{\epsilon},~\breve{\eta}). \label{infso}
\end{align}
\textcolor[rgb]{0.00,0.00,1.00}{where $\breve{W}$, $\breve{s}$, $\breve{\epsilon}$ and $\breve{\eta}$ denote the inferred weighted adjacency matrix, subconscious bias vector, confirmation bias parameter vector and novelty bias parameter vector that correspond to $[W]_{i,j} = w_{ij}$, $s = [s_{1},\ldots,s_{n}]^{\top}$, $\epsilon = [\epsilon_{1},\ldots,\epsilon_{n}]^{\top}$ and $\eta = [\eta_{1},\ldots,\eta_{n}]^{\top}$, respectively.}

\begin{figure}
\centering{
\includegraphics[scale=0.265]{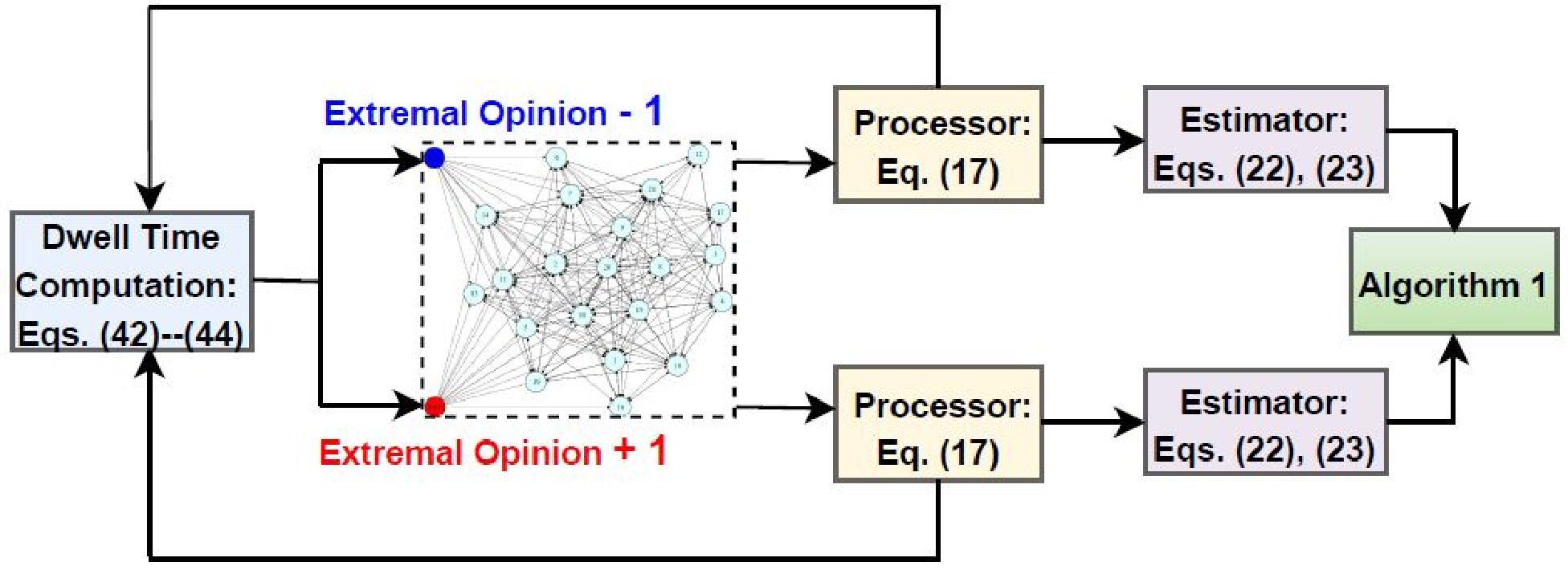}
}
\caption{Social system inference framework for solution \eqref{infso} with PAC: sample complexity determines the dwell time of strategic extremal opinions.}
\label{fig:exp}
\end{figure}

The proposed social system inference framework is first described in Figure \ref{fig:exp}, where the information sources are required to strategically and temporarily express extremal opinions $-1$ and $+1$ for obtaining the inference solution \eqref{infso}. This requirement is due to the state-dependent influence weights \eqref{eq: sdw1} and \eqref{sdw1}. If this requirement cannot be satisfied and is removed, i.e. the information sources have no cooperation, the information sources will be modeled as individuals in a social network, and the inference solution \eqref{infso} reduces  to $( {\breve{W},\breve{s}})$ for Friedkin-Johnsen model~\cite{friedkin1990social} or ${\breve{W}}$ for DeGroot model \cite{degroot1974reaching} as studied in \cite{wai2015social}.

Observing \eqref{eq: sdw1} and \eqref{sdw1}, we conclude that if information sources express extremal opinions $-1$ or $+1$, the social dynamics \eqref{kka} transforms to
a linear system, and if they express an identical opinion, they are regarded as one information source from the perspective of individuals. The formed linear systems under this consideration are described as follows.

\subsubsection{Extremal Opinion $-1$} Under the strategy that information sources express the identical extremal opinion $-1$, i.e., $h_{d}(k) = -1$ and $|\mathbb{K}| = 1$, the dynamics in \eqref{kka} transforms to
\begin{subequations}
\begin{align}
x( {k + 1}) &= \underline{\alpha}  + \underline{\mathcal{W}}x( k ) + \mathfrak{p}( k ) \label{sdw6aa}\\
y( {k} ) &= x(k) + \mathfrak{o}(k),\hspace{0.5cm}k \in \{1, \ldots, \underline{p}-1,\underline{p}\},\label{sdw6aaab}
\end{align}\label{sdw6}
\end{subequations}
where we define:
\begin{subequations}
\begin{align}
&{\underline{\alpha}_i} \triangleq ( {1 - \sum\limits_{j \in \mathbb{V}} {{w_{ij}}}} - \chi_{i})s_{i} - ( {{\epsilon_i} + {\eta _i}})( {1 + {s_i}}), \label{sdw72}\\
&{\left[ \underline{\mathcal{W}} \right]_{i,j}} \triangleq \begin{cases}
{w_{ii}} - \frac{{( {{s_i} + 1}){\eta _i}}}{{\sum\limits_{l \in \mathbb{V}} {{w_{il}}} }}{{{w_{ii}}}} + {( {1 + {s_i}}){\epsilon_i}}, \!&\text{if}~i \!=\! j \\
{w_{ij}} - \frac{{( {{s_i} + 1}){\eta _i}}}{{\sum\limits_{l \in \mathbb{V}} {{w_{il}}} }}{{{w_{ij}}}}, \!&\text{if}~i \!\neq\! j.
\end{cases}\label{sdw73}
\end{align}\label{sdw7}
\end{subequations}
\vspace{-0.10cm}
\subsubsection{Extremal Opinion $+1$} Under the strategy that information sources express the identical extremal opinion $+1$, i.e., $h_{d}(k) = +1$ and $|\mathbb{K}| = 1$, the dynamics in \eqref{kka} transforms to
\begin{subequations}
\begin{align}
x( {k + 1}) &= \overline{\alpha}  + \overline{\mathcal{W}}x(k) + \mathfrak{p}(k),   \label{sdw4a}\\
y( {k}) &= x(k) + \mathfrak{o}(k), \hspace{0.5cm}k \!\in\! \{\underline{p}\!+\!1, ~\ldots, ~\overline{p}\!-\!1, ~\overline{p}\},\label{sdw4aab}
\end{align}\label{sdw4}
\end{subequations}
where we define:
\begin{subequations}
\begin{align}
&{\overline{\alpha}_i} \triangleq ( {1 - \sum\limits_{j \in \mathbb{V}} {{w_{ij}}}} - \chi_{i}){s_i} + ( {{\epsilon_i} + \eta _i})( {1 - {s_i}}),\label{sdw52}\\
&{\left[ \overline{\mathcal{W}} \right]_{i,j}} \triangleq \begin{cases}
w_{ii} + \frac{{( {{s_i} - 1}){\eta _i}}}{{\sum\limits_{l \in \mathbb{V}} {{{{w_{il}}}}} }}{{{{w_{ii}}}}} + {( {1 - {s_i}}){\epsilon_i}}, \!&\text{if}~i\!=\! j \\
{w_{ij}} + \frac{{( {{s_i} - 1}){\eta _i}}}{{\sum\limits_{l \in \mathbb{V}} {{{{w_{il}}}}} }}{{{{w_{ij}}}}}, \!&\text{if}~i \!\neq\! j.
\end{cases}\label{sdw53}
\end{align}\label{sdw5}
\end{subequations}

\begin{rem}
Ignoring process noise $\mathfrak{p}( k )$, the dynamics \eqref{sdw6aa} and \eqref{sdw4a} have the same form of Friedkin-Johnsen model~\cite{friedkin1990social}, whose critical difference is that the matrix entries \eqref{sdw73} and \eqref{sdw53} of the dynamics \eqref{sdw6aa} and \eqref{sdw4a} are allowed to be negative, which is due to novelty bias. \label{cococo}
\end{rem}

To obtain \eqref{infso}, the social system inference framework first inputs observations of evolving opinions to estimate vectors $\overline{\alpha}$ and $\underline{\alpha}$ and matrices $\overline{\mathcal{W}}$ and $\underline{\mathcal{W}}$, which are denoted in
\begin{align}
\mathcal{M} \triangleq ( {\widehat{\overline{\alpha}},~~\widehat{\underline{\alpha}},~~\widehat{\overline{\mathcal{W}}},~~\widehat{\underline{\mathcal{W}}} }). \label{infmo}
\end{align}

\subsection{Problem Formulation}
\vspace{-0.10cm}
In the context of learning-based competing camps in social networks, information sources leverage inference solution \eqref{infso} for later decision making  of optimal information spread strategy. Therefore, information sources can only express extremal opinions \textcolor[rgb]{0.00,0.00,1.00}{temporarily} to infer \eqref{infso}. Hereto, considering \eqref{sdw6aaab} and \eqref{sdw4aab}, we introduce the dwell times of  strategic extremal opinions $-1$ and $+1$, respectively, as
\begin{align}
\textcolor[rgb]{0.00,0.00,1.00}{\tau_{-1} \triangleq \underline{p} \!-\! 1 \!+\! 1 \!=\! \underline{p}, ~~~~\tau_{+1} \triangleq \overline{p} \!-\! (\underline{p}+1) \!+\! 1 \!=\! \overline{p} \!-\! \underline{p}.  \label{cjkd}}
\end{align}
The computation of dwell times that are sufficient for high accurate inference \eqref{infso} constitutes the first problem of this paper. Observing \eqref{sdw7} and \eqref{sdw5}, we discover that the matrices $\underline{\mathcal{W}}$ and $\overline{\mathcal{W}}$ include all of the inference objectives included in \eqref{infso}. Thus, the dwell times are determined by  $\overline{\mathcal{W}}$ and $\underline{\mathcal{W}}$ estimations in this problem, which is formally stated as follows.

\begin{prom}  Find the dwell times of strategic extremal opinions that guarantee the estimations of $\overline{\mathcal{W}}$ and $\underline{\mathcal{W}}$ to be $(\phi, \delta)$--PAC, i.e.
\begin{align}
\!\!\mathbf{P} (||\widehat{\overline{\mathcal{W}}}  \!-\! {\overline{\mathcal{W}}} || \leq \phi) \!\geq\! 1 \!-\! \delta ~\text{and}~\mathbf{P} (|| \widehat{\underline{\mathcal{W}}} \!-\! {\underline{\mathcal{W}}} || \!\leq\! \phi) \!\geq\! 1 \!-\! \delta \label{pac}
\end{align}
for $\phi > 0$ and $0 < \delta < 1$. \label{prom1}
\end{prom}

Given the estimation solution \eqref{infmo}, inferring \eqref{infso} constitutes the second problem.
\begin{prom} Determine a social system inference procedure that generates the inference solution \eqref{infso}. \label{prom2}
\end{prom}

\section{Problem \ref{prom1}: Social System Estimation}
In this section, we first present the least-square auto-regressive estimation. We then present assumptions and investigate sample complexity of estimation, \textcolor[rgb]{0.00,0.00,1.00}{leveraging which we derive the dwell times of strategic extreme opinions to guarantee the inference solution \eqref{infso} to achieve $\left(\phi, \delta\right)$--PAC}. To simplify the representations of investigations, we define:
\begin{subequations}
\begin{align}
\vartheta(k)\footnotemarkmath &\!\triangleq\! \begin{cases}
-1, &k \!\in\! \{1, \ldots, \underline{p}\}\\
+1, &k \!\in\! \{\underline{p} + 1, \ldots, \overline{p}\}
\end{cases} \\
\mathcal{A}_{\vartheta} &\!\triangleq\! \begin{cases}
\overline{\mathcal{W}}, &\vartheta \!=\! +1 \\
\underline{\mathcal{W}},&\vartheta \!=\! -1
\end{cases}~~~~~~~~~~~~~\mathfrak{a}_{\vartheta} \!\triangleq\! \begin{cases}
\overline{\alpha}, &\vartheta \!=\! +1 \\
\underline{\alpha},&\vartheta \!=\! -1
\end{cases}\!\label{pvp}\\
k_{\vartheta} &\!\triangleq\! \begin{cases}
\underline{p} + 1, &\vartheta \!=\! +1 \\
1,&\vartheta \!=\! -1
\end{cases}~~~~~~~~~p_{\vartheta} \!\triangleq\! \begin{cases}
\overline{p}, &\vartheta \!=\! +1 \\
\underline{p}, &\vartheta \!=\! -1
\end{cases}
\end{align}\label{mmck}
\end{subequations}
\footnotetextmath{In this paper, we use $\vartheta(k)$ and $\vartheta$ interchangeably.}
\!\!\!\!\!based on which, we obtain the trajectory of observed public opinions from the dynamics \eqref{sdw6} and \eqref{sdw4} as
\begin{align}
y(j) \!=\! \sum\limits_{i = 1}^{j - 1}\! {{\mathcal{M}_{(i,j)}}( {{\textcolor[rgb]{0.00,0.07,1.00}{\mathfrak{a}_{\vartheta(j-i)}}} \!+\! \mathfrak{p}(j\!-\!i)}) \!+\! \mathcal{M}_{j}^\mathrm{c}}x(1)  \!+\! \mathfrak{o}(j), \label{traa}
\end{align}
where
\begin{subequations}
\begin{align}
{\mathcal{M}_{( {i,j})}} &\triangleq \begin{cases}
\mathcal{A}_{+1}^{j-\underline{p}}\!\mathcal{A}_{-1}^{i - j + \underline{p} - 1}, &j > \underline{p}~\text{and}~i > j-\underline{p} \\
\mathcal{A}_{+1}^{i-1}, &j > \underline{p}~\text{and}~i \leq j-\underline{p} \\
\mathcal{A}_{-1}^{i-1}, &j \le \underline{p}
\end{cases}, \\
\mathcal{M}^{\mathrm{c}}_{j} &\triangleq \begin{cases}
\mathcal{A}_{+1}^{j - \underline{p}}\mathcal{A}_{-1}^{\underline{p} - 1}, &j > \underline{p} \\
\mathcal{A}_{-1}^{j - 1}, &j \le \underline{p}.
\end{cases}
\end{align}\label{traaab}
\end{subequations}

\subsection{Data Processor}
\vspace{-0.10cm}
We now present data processor of observations of public evolving  opinions, as shown in Figure \ref{fig:exp}, which is a necessary step for the sample complexity analysis.
\begin{align}
\widetilde{\mathbf{y}}^{j}_{g} \triangleq y(g) - y(j), ~~~g < j \in \{1,2,\ldots,\overline{p}\}. \label{dp1}
\end{align}
Correspondingly, we define:
\begin{align}
\widetilde{\mathbf{x}}^{j}_{g} \!\triangleq\! x(g) \!-\! x(j), ~~
\widehat{\mathfrak{p}}^{j}_{g} \!\triangleq\! \mathfrak{p}(g) \!-\! \mathfrak{p}(j), ~~
\widetilde{\mathfrak{o}}^{j}_{g} \!\triangleq\! \mathfrak{o}(g) \!-\! \mathfrak{o}(j).\label{ppqq}
\end{align}
With the consideration of \eqref{mmck}, \eqref{dp1} and \eqref{ppqq}, we obtain the following dynamics from systems \eqref{sdw6} and \eqref{sdw4}:
\begin{align}
\widetilde{\mathbf{x}}^{j+1}_{g+1} = \mathcal{A}_{\vartheta}\widetilde{\mathbf{x}}^{j}_{g} + \widehat{\mathfrak{p}}^{j}_{g}, ~~~\widetilde{\mathbf{y}}^{j}_{g} = \widetilde{\mathbf{x}}^{j}_{g} + \widetilde{\mathfrak{o}}^{j}_{g}, ~~g\!<\! j \!\leq\! \underline{p}~\text{or}~j \!>\! g \!>\! \underline{p}\nonumber
\end{align}
by which we then obtain the dynamics of $\widetilde{\mathbf{y}}^{j}_{g}$ as
\vspace{-0.1cm}
\begin{subequations}
\begin{align}
\widetilde{\mathbf{y}}^{j+1}_{g+1} &= \mathcal{A}_{\vartheta}\widetilde{\mathbf{y}}^{j}_{g} + \widehat{\mathfrak{g}}^{j}_{g} \\
\widehat{\mathfrak{g}}^{j}_{g} &= \widehat{\mathfrak{p}}^{j}_{g} + \widetilde{\mathfrak{o}}^{j+1}_{g+1} - \mathcal{A}_{\vartheta}\widetilde{\mathfrak{o}}^{j}_{g},\!~~g\!<\! j \!\leq\! \underline{p}~\text{or}~j \!>\! g \!>\! \underline{p}.\label{zgg1re}
\end{align}\label{zgg1}
\end{subequations}
\vspace{-0.50cm}
\begin{rem}
The relation \eqref{zgg1re} explicitly shows the statistical dependence of random vector $\widehat{\mathfrak{g}}^{j}_{g}$ on time indexed by $j$ and $g$. Meanwhile, \eqref{zgg1re} also indicates the statistical dependence of $\widehat{\mathfrak{g}}^{j}_{g}$ on its coordinates, \textcolor[rgb]{0.00,0.00,1.00}{i.e., the covariance matrix of $\widehat{\mathfrak{g}}^{j}_{g}$ is not a diagonal matrix, which is due to the term $\mathcal{A}_{\vartheta}\widetilde{\mathfrak{o}}^{j}_{g}$}.
\end{rem}

\vspace{-0.39cm}
\subsection{Model Estimation}
\vspace{-0.10cm}
We construct the following data matrices:
\begin{align}
&Y_{\vartheta} \!\triangleq\! \left[\widetilde{\mathbf{y}}^{k_{\vartheta}+2}_{k_{\vartheta}+1},\ldots, \widetilde{\mathbf{y}}^{p_{\vartheta}}_{k_{\vartheta}+1}, \widetilde{\mathbf{y}}^{k_{\vartheta}+3}_{k_{\vartheta}+2}, \ldots, \widetilde{\mathbf{y}}^{p_{\vartheta}}_{k_{\vartheta}+2}, \ldots, \widetilde{\mathbf{y}}^{p_{\vartheta}}_{p_{\vartheta}-1}\right],\nonumber\\
&U_{\vartheta} \!\triangleq\! \left[\widehat{\mathfrak{g}}^{k_{\vartheta}+1}_{k_{\vartheta}},\ldots, \widehat{\mathfrak{g}}^{p_{\vartheta}-1}_{k_{\vartheta}}, \widehat{\mathfrak{g}}^{k_{\vartheta}+2}_{k_{\vartheta}+1}, \ldots,  \widehat{\mathfrak{g}}^{p_{\vartheta}-1}_{k_{\vartheta}+1}, \ldots, \widehat{\mathfrak{g}}^{p_{\vartheta}-1}_{p_{\vartheta}-2}\right],\nonumber\\
&X_{\vartheta} \!\triangleq\! \left[\widetilde{\mathbf{y}}^{k_{\vartheta}\!+\!1}_{k_{\vartheta}},\ldots, \widetilde{\mathbf{y}}^{p_{\vartheta}\!-\!1}_{k_{\vartheta}}, \widetilde{\mathbf{y}}^{k_{\vartheta}\!+\!2}_{k_{\vartheta}\!+\!1}, \ldots, \widetilde{\mathbf{y}}^{p_{\vartheta}\!-\!1}_{k_{\vartheta}\!+\!1}, \ldots, \widetilde{\mathbf{y}}^{p_{\vartheta}\!-\!1}_{p_{\vartheta}\!-\!2}\right],\label{xf211}
\end{align}
considering which, we verify from system \eqref{zgg1} that
\begin{align}
{Y}_{\vartheta} = \mathcal{A}_{\vartheta}{X}_{\vartheta} + {{U}}_{\vartheta}.   \label{datanew11}
\end{align}
We note that matrix ${{U}}_{\vartheta}$ is unknown. The relation \eqref{zgg1} thus indicates the least-square optimal estimation of $\mathcal{A}_{\vartheta}$ is
\begin{align}
{\widehat{\mathcal{A}}_{\vartheta}} = {Y}_{\vartheta}{X}^{\top}_{\vartheta}{({X}_{\vartheta}{X}^{\top}_{\vartheta})^{ - 1}}. \label{esm}
\end{align}
Considering \eqref{sdw6}, \eqref{sdw4} and \eqref{pvp}, with obtained estimation \eqref{esm}, the estimation of $\mathfrak{a}_{\vartheta(k)}$ is
\begin{align}
\widehat{\mathfrak{a}}_{\vartheta} = \frac{1}{{p_{\vartheta} - k_{\vartheta}}}\sum\limits_{q = k_{\vartheta}}^{p_{\vartheta} - 1} {( y( {q + 1}) - \widehat{\mathcal{A}}_{\vartheta}y( q))}.\label{esmok}
\end{align}

Finally, we denote two matrices according to \eqref{xf211}:
\begin{align}
\mathbf{E}[ {{X_{\vartheta}}X_{\vartheta}^\top}] = \sum\limits_{g = k_{\vartheta}}^{p_{\vartheta} - 2} {\sum\limits_{g < j = k_{\vartheta} + 1}^{p_{\vartheta} - 1}\!\!\! \textcolor[rgb]{0.00,0.00,1.00}{\mathbf{E}[{\widetilde{\mathbf{y}}_g^j{{( {\widetilde{\mathbf{y}}_g^j})^\top}}}}]}  &\triangleq {\Gamma _{\vartheta}} \triangleq \Psi^{-2}_{\vartheta}. \label{oEp2}
\end{align}

\subsection{Assumption}
We construct the following stacked vectors and matrices for presenting assumptions on answering Problem \ref{prom1}:
\begin{align}
&{\theta}_{\vartheta} \!\triangleq\! [{{\mathbf{o}}_{\vartheta}}; {{\mathbf{p}}_{\vartheta}}],~{\widehat{\theta}}_{\vartheta} \!\triangleq\! [{{\mathbf{a}}_{\vartheta}};\mathbf{x}_{\vartheta}],~{\overline{\theta}_{\vartheta}} \!\triangleq\!  {[ {\theta _{\vartheta};{\mathbf{0}}}]},~{\underline{\theta}_{\vartheta}} \!\triangleq\!  {[ {{\mathbf{0}};\widehat{\theta}_{\vartheta}}]},\label{tr1aas}\\
&{{\Upsilon}_{\vartheta}} \triangleq \text{diag}\{\!{\Psi _{\vartheta}}, ~\Psi_{\vartheta},~\ldots, ~{\Psi _{\vartheta}} \!\}  \!\in\! \mathbb{R}^{\mathfrak{l}_{\vartheta} \times \mathfrak{l}_{\vartheta}}, \label{adef1}\\
&\mathbf{U} \triangleq \text{diag}\left\{ {\mathbf{u}},~{\mathbf{u}}, ~\ldots,~{\mathbf{u}} \right\} \!\in\! \mathbb{R}^{\mathfrak{l}_{\vartheta} \times \frac{\mathfrak{l}_{\vartheta}}{n}}, \label{adef1kk}\\
&{\Pi_{\vartheta}} \triangleq [{\mathbf{I}_{\mathfrak{l}_{\vartheta}},~{\overline{\mathrm{W}}}_{\vartheta},~{\overline{\mathrm{W}}}_{\vartheta},~{\breve{\overline{\mathrm{W}}}}_{\vartheta}}], \label{gh9}
\end{align}
where ${{\mathbf{o}}_{\vartheta}}$, ${\mathbf{p}}_{\vartheta}$, ${{\mathbf{a}}_{\vartheta}}$, $\mathbf{x}_{\vartheta}$, ${\breve{\overline{\mathrm{W}}}}_{\vartheta}$ and ${\overline{\mathrm{W}}}_{\vartheta}$ are defined by \eqref{asv1}-\eqref{ghmm2} in Appendix A, $\mathbf{u} \in {\mathcal{S}^{n - 1}}$, $\Psi _{\vartheta}$ is defined in \eqref{oEp2}, and
\begin{align}
{\mathfrak{l}_{\vartheta}} \triangleq \frac{{1}}{2}(p_{\vartheta} - k_{\vartheta})(p_{\vartheta} - k_{\vartheta} - 1)n, \label{ghcc}
\end{align}

\textcolor[rgb]{0.00,0.00,1.00}{Using the same observation vectors $\widetilde{\mathbf{y}}^{j}_{g}$, $g < j \in \{k_{\vartheta}, \ldots, p_{\vartheta}-1\}$ in constructing the data matrix \eqref{xf211}, we construct a data vector:}
\begin{align}
\!\!\!\breve{\mathrm{x}}_{\vartheta} \!\triangleq\!\! \left[ {\widetilde{\mathbf{y}}_{{k_{\vartheta}}}^{{k_{\vartheta}} + 1}; \ldots; \widetilde{\mathbf{y}}_{{k_{\vartheta}}}^{{p_{\vartheta}} - 1}; \widetilde{\mathbf{y}}_{{k_{\vartheta}} + 1}^{{k_{\vartheta}} + 2}; \ldots ; \widetilde{\mathbf{y}}_{{k_{\vartheta}} + 1}^{{p_{\vartheta}} - 1}; \ldots ;\widetilde{\mathbf{y}}_{{p_{\vartheta}} - 2}^{{p_{\vartheta}} - 1}} \right]\!\!, \label{datavector}
\end{align}
by which we verify from \eqref{tr1aas},  \eqref{gh9} and \eqref{ghcc} that
\begin{align}
\breve{\mathrm{x}}_{\vartheta} = {\Pi_{\vartheta}}({\overline{\theta}_{\vartheta}}+{\underline{\theta}_{\vartheta}}). \label{ghhbbc}
\end{align}

To introduce the assumption setting for our estimation and inference, we now recall a definition: \textcolor[rgb]{0.00,0.07,1.00}{\begin{definition}[Convex Concentration Property \cite{ledoux2001concentration}] Let $\mathbf{x}$ be a random vector in $\mathbb{R}^{n}$. Then $\mathbf{x}$ has the
convex concentration property with constant $\kappa$, if for every 1-Lipschitz convex function $\varphi: \mathbb{R}^{n} \rightarrow \mathbb{R}$, we have $\mathbf{E}\left[ {\left| {\varphi(\mathbf{x})} \right|} \right] < \infty$ and for every time $t > 0$, we have
\begin{align}
\mathbf{P}\left[ {\left| {\varphi(x) - \mathbf{E}[ {\varphi ( x)}]} \right| \geq t} \right] \leqslant 2{e^{ - \frac{{{t^2}}}{{{\kappa ^2}}}}}.
\end{align}
\end{definition}}

With the definitions at hand, we make the following assumptions for solving Problem \ref{prom1}.
\begin{asm}
Consider the expectation \eqref{oEp2} and the social dynamics \eqref{kka} with \eqref{zgg1re} and \eqref{tr1aas}--\eqref{gh9}.
\begin{enumerate}
  \item $\mathfrak{p}(k)$ $\iidsim$ $\mathcal{D}_{\mathrm{p}}(\mathbf{0},\sigma^{2}_{\mathrm{p}}\mathbf{I}_{n})$, ~~$\mathfrak{o}(k)$ $\iidsim$ $\mathcal{D}_{\mathrm{o}}(\mu_{\mathrm{o}}\mathbf{1},\sigma^{2}_{\mathrm{o}}\mathbf{I}_{n})$.
  \item ${\overline{\theta}}_{\vartheta}$ has the convex concentration property with constant $\kappa > 0$.
  \item $[\widehat{\mathfrak{g}}^{j}_{k}]_{i}$, $i \in \mathbb{V}$, \textcolor[rgb]{0.00,0.07,1.00}{is $\mathcal{F}_{k}$-measurable (i.e., measurable with respect to the filtration $\mathcal{F}_{k} \triangleq \sigma([\widehat{\mathfrak{g}}^{j}_{s}]_{i}, s \leq k)$)} and conditionally $\gamma$-sub-Gaussian for some $\gamma$ $>$ $0$, i.e., $\mathbf{E}\!\left[ {\left. {{e^{\lambda \left[ {\widehat{\mathfrak{g}}_{k+1}^{j + 1}} \right]}}} \right|{\mathcal{F}_k}} \right]$, for all $\lambda$ \!$\in$\! $\mathbb{R}$, $k \!<\! j \!\in\! \{k+1,\ldots,p-1\}$.
  \item There exist scalars $\varrho_{1} > 0$ and $\varrho_{2} > 0$, such that
\begin{align}
&\left| || \mathbf{U}^{\top}{{\Upsilon}^{\top}_{\vartheta}}{{\Pi}_{\vartheta}}{{{\overline{\theta}}_{\vartheta}}} ||_2^2 + || \mathbf{U}^{\top}{{\Upsilon}^{\top}_{\vartheta}}{{\Pi}_{\vartheta}}{{{\underline{\theta}}_{\vartheta}}} ||_2^2 - 1 \right| \nonumber\\
&\geq \left| \varrho_{1} -  \varrho_{2}||\mathbf{U}^{\top}{{\Upsilon}^{\top}_{\vartheta}}{{\Pi}_{\vartheta}}({\overline{\theta}}_{\vartheta} + {\underline{\theta}}_{\vartheta}) ||_2^2\right|. \nonumber
\end{align}
\end{enumerate}\label{asprv}
\end{asm}

\begin{rem}
\textcolor[rgb]{0.00,0.00,1.00}{The subscripts $\mathrm{p}$ and $\mathrm{o}$ in Assumptions \ref{asprv}-1) are used to indicate that the distributions of process noise $\mathfrak{p}(k)$ and observation noise $\mathfrak{o}(k)$ can be different.} Examples under Assumption \ref{asprv}-2), as summarized in \cite{adamczak2015note}, include any random vector $\eta \in \mathbb{R}^{s}$ with independent coordinates and almost sure $|[\eta]_{i}| \leq 1$ for any $i \in \{1, \ldots, s\}$, \textcolor[rgb]{0.00,0.00,1.00}{random vectors obtained via sampling without replacement \cite{paulin2014convex}}, vectors with bounded coordinates satisfying some uniform mixing conditions or Dobrushin type criteria. Examples of $[\widehat{\mathfrak{g}}^{j}_{k}]_{i}$ under Assumption \ref{asprv}-3) include a bounded zero-mean noise lying in an interval of length at most $2\gamma$, a zero-mean Gaussian noise with variance at most $\gamma^{2}$ \cite{abbasi2011improved}. \textcolor[rgb]{0.00,0.00,1.00}{Under Assumption \ref{asprv}-2), Lemma \ref{newr} in Appendix B is employed to derive \eqref{mcp1} and \eqref{mcp2} in Appendix C. Assumption \ref{asprv}-4) is leveraged to derive \eqref{relatmc2} in Appendix C. }
\end{rem}

\begin{rem}[Non-Zero Mean] If $\mathfrak{p}(k)$ $\iidsim$ $\mathcal{D}_{\mathrm{p}}(\mu_{\mathrm{p}}\mathbf{1},\sigma^{2}_{\mathrm{p}}\mathbf{I}_{n})$, it can be rewritten as $\mathfrak{p}(k) = \mu_{\mathrm{p}}\mathbf{1} + \widetilde{\mathfrak{p}}(k)$, with $\widetilde{\mathfrak{p}}(k)$ $\iidsim$ $\mathcal{D}_{\mathrm{p}}(\mathbf{0},\sigma^{2}_{\mathrm{p}}\mathbf{I}_{n})$. In this scenario, model \eqref{sdw4a}, as an example, can be written as $x( {k + 1}) = (\overline{\alpha} + \mu_{\mathrm{p}}\mathbf{1}) + \overline{\mathcal{W}}x(k) + \widetilde{\mathfrak{p}}(k)$. Thus,  Assumption \ref{asprv}-1) on process noise holds in general.
\end{rem}

\subsection{Sample Complexity}
\vspace{-0.00cm}
We now investigate the sample complexity of estimation \eqref{esm}, whose associated conditions will answer Problem \ref{prom1}.

Under Assumption \ref{asprv}-1), we obtain the covariance matrix of vector ${\theta}_{\vartheta}$ given in \eqref{tr1aas} as
\begin{align}
{\mathcal{C}}_{\vartheta} \triangleq {\bf{E}}\left[{\theta}_{\vartheta}{\theta}^{\top}_{\vartheta}\right] = \left[\! {\begin{array}{*{20}{c}}
2{\sigma_{\mathrm{o}}^2{\mathbf{I}_{{\mathfrak{l}_{\vartheta}}}}}&\mathbf{O}\\
\mathbf{O}&{\bf{E}}[{{\mathbf{p}}_{\vartheta}}{{\mathbf{p}}^{\top}_{\vartheta}}]\footnotemarkmath
\end{array}} \!\right],\label{kop2ab}
\end{align}\footnotetextmath{With \eqref{ghb6}--\eqref{ghb4}, ${\bf{E}}[{{\mathbf{p}}_{\vartheta}}{{\mathbf{p}}^{\top}_{\vartheta}}]$ can be straightforwardly computed under Assumption \ref{asprv}-1), which is not presented in this paper due to page limit.}
\!\!\!where $\mathbf{O}$ denotes zero matrix with compatible dimensions. Considering the covariance matrix, we present an auxiliary proposition, whose proof appears in Appendix C.
\begin{prop}
Under Assumption \ref{asprv}, we have
\begin{align}
&\mathbf{P}\left[ {|| \varrho_{2}\Psi_{\vartheta}^\top{X_{\vartheta}}X_{\vartheta}^\top\!{\Psi_{\vartheta}} - \varrho_{1}\mathbf{I}_{n}|| > \rho }\right] \nonumber\\
&\leq 2 \cdot {( {\frac{2}{\varepsilon } + 1})^n} \cdot e^{\frac{-1}{{\mathfrak{c}{\kappa^2}}}\min \left\{ {\frac{{{(1 - 2\varepsilon )^2n\rho ^2}}}{{\mathfrak{l}_{\vartheta}||{{\Pi}^{\top}_{\vartheta}}\!{\Upsilon}_{\vartheta}||^{4}|| {\mathcal{C}}_{\vartheta} ||}},~\frac{{(1 - 2\varepsilon)\rho}}{{||{{\Pi}^{\top}_{\vartheta}}{{\Upsilon}_{\vartheta}} ||^{2}}}}\right\}}\label{ncffn}
\end{align}
for $\varepsilon \in [0, \frac{1}{2})$ and some universal constant $\mathfrak{c} > 0$. \label{ptopk1a}
\end{prop}

Leveraging Proposition \ref{ptopk1a}, the sample complexity is presented in the following theorem, whose proof is presented in Appendix D.
\begin{thm} Consider the estimated matrix ${\widehat{\mathcal{A}}_{\vartheta}}$ in \eqref{esm}, and the real matrix $\mathcal{A}_{\vartheta}$ in \eqref{mmck}. Under Assumption \ref{asprv}, for any \textcolor[rgb]{0.00,0.07,1.00}{$\varepsilon \in [0, \frac{1}{2})$, $\rho \in (0, \varrho_{1})$, $\delta \in (0,1)$, $\overline{\varepsilon} \in [0,1)$ and $\phi > 0$}, we have
\begin{align}
\mathbf{P}[||{\widehat{\mathcal{A}}_{\vartheta}} - \mathcal{A}_{\vartheta}|| > \phi] \leq \delta\label{fnfn}
\end{align}
if the \textcolor[rgb]{0.00,0.00,1.00}{following} hold:
\begin{align}
&\min\! \left\{\!{\frac{{{{(1 \!-\! 2\varepsilon )}^2}n{\rho ^2}}}{{\mathfrak{l}_{\theta}||\Pi _{\vartheta}^\top\! {\Upsilon _{\vartheta}}||^4||{\mathcal{C}_{\vartheta}}||}},\frac{{(1 \!-\! 2\varepsilon )\rho }}{{||\Pi_{\vartheta}^\top\!{\Upsilon _{\vartheta}}||^2}}}\!\right\} \!\!\ge\! \frac{{\gamma ^2}}{2} \!\ln \!\frac{{4 {( {\frac{2}{\varepsilon } \!+\! 1})^n}}}{\delta }, \label{nc1}\\
&{\lambda _{\min }}( {{\Gamma _{\vartheta}}}) \!\ge\! \frac{{32\mathfrak{c}{\kappa^2}\varrho_{2}}}{{{\phi^2}(\varrho_{1} \!-\! \rho)}}\! \ln \!\!\left(\!\frac{{2 \!\cdot\! (2 \!+\! \overline{\varepsilon})^{n}}}{\delta\cdot\overline{\varepsilon}^{n}}{{\left(\! {\frac{{2\varrho_{1}}}{{\varrho_{1} \!-\! \rho}}} \!\right)\!}^{0.5n}}\!\right) \textcolor[rgb]{0.00,0.00,1.00}{\!>\! 0}. \label{nc2}
\end{align}\label{ptopk1er}
\end{thm}
\vspace{-0.20cm}
\begin{rem}
The dwell times of strategic extremal opinions can be computed from the conditions \eqref{nc1} and \eqref{nc2}. However, the current forms are not ready for the computation, which is due to the unknown $\mathcal{A}_{\vartheta}$  included in $\Pi_{\vartheta}$ and ${\Upsilon _{\vartheta}}$. \textcolor[rgb]{0.00,0.00,1.00}{With the consideration of $x_{i}(k) \in [-1,1]$, $\forall i \in \mathbb{V}$, $\forall k \in \mathbb{N}$, the subsystems \eqref{sdw6} and \eqref{sdw4} indicate that the switching matrix $\mathcal{A}_{\vartheta}$ defined in \eqref{pvp} is Schur stable. Therefore, it is practical to assume that we know matrix-norm bounds $\underline{\mathfrak{h}}$ and $\overline{\mathfrak{h}}$ such that $0 < \underline{\mathfrak{h}} \leq \left\| \mathcal{A}_{\vartheta} \right\| \leq \overline{\mathfrak{h}} \leq 1$. The bounds $\underline{\mathfrak{h}}$ and $\overline{\mathfrak{h}}$ can be leveraged to estimate the bounds on the matrix norms of $\Pi_{\vartheta}$ and ${\Upsilon _{\vartheta}}$ to compute the dwell times, which will be carried out in next subsection.}
\end{rem}

\vspace{-0.20cm}
\subsection{Dwell Times of Strategic Extremal Opinions}
With the consideration of $\mathcal{A}_{\vartheta}$ in \eqref{mmck} and $\Pi_{\vartheta}$ in \eqref{gh9}, we present the following bounds pertaining to matrix norm.
\begin{align}
&\mathop {\min }\limits_{\!\!i,j,g,h \in \mathbb{N}_{0}}\!\! \left\{\! {{{|| \mathcal{A}_{+1}^{j}\!\mathcal{A}_{-1}^{i}\!||}^{2}\!, {|| \mathcal{A}_{+1}^{g}\!\mathcal{A}_{-1}^{h}\!(\mathbf{I}_{n} \!-\! \mathcal{A}_{+1}^{j}\!\mathcal{A}_{-1}^{i}\!)||}^{2}}} \!\right\} \!\geq\! {\underline{\mathfrak{s}}}, \label{nc1dab}\\
&\mathop {\max }\limits_{k < p \in \mathbb{N}_{0}} \left\{ {{{|| \Pi_{(k,p)}||^{2}}}} \right\} \leq {\overline{\mathfrak{s}}} \leq 1. \label{nc1dcc}
\end{align}

\begin{rem}
The inequality \eqref{nc1dcc} is obtained via considering \eqref{ghhbbc}, where $-\mathbf{1} \leq \breve{\mathrm{x}}_{\vartheta} \leq \mathbf{1}$ and $-\mathbf{1} \leq {\overline{\theta}_{\vartheta}}+{\underline{\theta}_{\vartheta}} \leq \mathbf{1}$ hold for any $k < p \in \mathbb{N}$ and any $x_{i}(1) \in [-1,1], i \in \mathbb{V}$.
\end{rem}

With the bounds given in \eqref{nc1dab} and \eqref{nc1dcc}, we define:
\begin{align}
\mathfrak{f}_{\vartheta} &\triangleq 2\frac{{\mathfrak{l}_{\vartheta}}}{n}\sigma^{2}_{\mathrm{o}} + \underline{\mathfrak{s}}\frac{{\textcolor[rgb]{0.00,0.00,1.00}{\overline{\mathfrak{l}}_{\vartheta}}}}{n}\sigma^{2}_{\mathrm{p}}, ~~~~~~~~~{\mathfrak{j}_{\vartheta}} \triangleq \frac{\overline{\mathfrak{s}}}{\mathfrak{f}_{\vartheta}},\label{nad1tt} \\
\textcolor[rgb]{0.00,0.07,1.00}{{\overline{\mathfrak{l}}_{\vartheta}}} &\triangleq  \sum\limits_{i = 1}^{p_{\vartheta} - k_{\vartheta} - 1} {( {p_{\vartheta} - k_{\vartheta} - i})( {p_{\vartheta} - 1 - i})n}. \label{ghccan}
\end{align}

With the definitions at hand, we present a corollary of Theorem \ref{ptopk1er}, whose proof is given in Appendix E.
\begin{cor}
The conditions \eqref{nc1} and \eqref{nc2} hold if
\begin{align}
&\min \left\{ {\frac{{{{(1 \!-\! 2\varepsilon )}^2}n{\rho ^2}}}{{\mathfrak{l}_{\vartheta}{\mathfrak{j}^{2}_{\vartheta}}||{\mathcal{C}_{\vartheta}}||}},~\frac{{(1 \!-\! 2\varepsilon )\rho }}{{{\mathfrak{j}_{\vartheta}}}}} \right\} \ge \frac{{\gamma ^2}}{2}\ln \frac{{4 {( {\frac{2}{\varepsilon } + 1})^n}}}{\delta}, \label{fccq}\\
&\mathfrak{f}_{\vartheta} \geq \frac{{32\mathfrak{c}{\kappa^2}\varrho_{2}}}{{{\phi^2}(\varrho_{1} \!-\! \rho)}}\! \ln \!\left(\frac{{2 \!\cdot\! (2 \!+\! \overline{\varepsilon})^{n}}}{\delta\cdot\overline{\varepsilon}^{n}}{{\left( {\frac{{2\varrho_{1}}}{{\varrho_{1} \!-\! \rho}}} \right)}^{0.5n}}\right) > 0.\label{fcc1}
\end{align}
\label{fth}
\end{cor}
\vspace{-0.20cm}
\begin{rem}[Dwell Time Computation]
According to \eqref{cjkd} and Corollary \ref{fth}, dwell time is computed as
\begin{align}
\tau_{\vartheta} = p_{\vartheta} - k_{\vartheta} + 1, ~s.t. ~\eqref{nad1tt}-\eqref{fcc1}. \label{cook}
\end{align}
\end{rem}

\vspace{-0.50cm}
\textcolor[rgb]{0.00,0.00,1.00}{\begin{rem}
The conditions \eqref{fccq} and \eqref{fcc1} straightforwardly indicate that they are more likely to hold for larger $\delta$ or $\phi$, i.e., the smaller prescribed level of accuracy or confidence, which can lead to smaller dwell time implied by \eqref{ghccan}. The conditions \eqref{fccq} and \eqref{fcc1} also imply that given the dwell times, the larger size of social network $n$ can require larger $\delta$ or $\phi$, which can further result in larger model error. The condition \eqref{fcc1} and the definition \eqref{ghccan} imply that given $\delta$ or $\phi$, the smaller $\kappa$ and $\gamma$ can also result in the smaller dwell time.
\end{rem}}

\vspace{-0.20cm}
\section{Problem \ref{prom2}: Social System Inference}
\vspace{-0.10cm}
With the obtained estimation \eqref{infmo}, we investigate the computation of \eqref{infso}. Considering the structures of real vectors and matrices in \eqref{sdw7} and \eqref{sdw5}, we write the estimations \eqref{esm} and \eqref{esmok} in the following forms:
\begin{subequations}
\begin{align}
{[\widehat{\mathcal{A}}_{+1}]_{i,j}} &=  \begin{cases}
\!{{\breve{w}_{ii}} + \frac{{( {{\breve{s}_i} - 1}){\breve{\eta}_i}}}{{\sum\limits_{l \in \mathbb{V}} {{{{{\breve{w}_{il}}}}}} }}{{{{\breve{w}_{ii}}}}}+ ( {1 - {\breve{s}_i}}){\breve{\epsilon}_i}} &\text{if}~i \!=\! j \\
\!{\breve{w}_{ij}} + \frac{{( {{\breve{s}_i} - 1}){\breve{\eta}_i}}}{{\sum\limits_{l \in \mathbb{V}} {{{{{\breve{w}_{il}}}}}} }}{{{{\breve{w}_{ij}}}}},&\text{if}~i \!\neq\! j
\end{cases}\label{dec2a}\\
{[\widehat{\mathcal{A}}_{-1}]_{i,j}} &=  \begin{cases}
\!{{\breve{w}_{ii}} - \frac{{( {{\breve{s}_i} + 1}){\breve{\eta}_i}}}{{\sum\limits_{l \in \mathbb{V}} {{{{{\breve{w}_{il}}}}}} }}{{{{\breve{w}_{ii}}}}}  + ( {1 + {\breve{s}_i}}){\breve{\epsilon}_i}} &\text{if}~i \!=\! j \\
\!{\breve{w}_{ij}} - \frac{{( {{\breve{s}_i} + 1}){\breve{\eta}_i}}}{{\sum\limits_{l \in \mathbb{V}} {{{{{\breve{w}_{il}}}}}} }}{{{{\breve{w}_{ij}}}}},&\text{if}~i \!\neq\! j
\end{cases}\label{dec2b}\\
[\widehat{\mathfrak{a}}_{+1}]_{i} &= ( {1 - \sum\limits_{j \in \mathbb{V}} {{\breve{w}_{ij}}}}){\breve{s}_i} + ( {{\breve{\epsilon}_i} + \breve{\eta}_i})( {1 - {\breve{s}_i}}),\label{dec2b1}\\
[\widehat{\mathfrak{a}}_{-1}]_{i} &= ( {1 - \sum\limits_{j \in \mathbb{V}} {{\breve{w}_{ij}}}})\breve{s}_{i} - ( {{\breve{\epsilon}_i} + {\breve{\eta}_i}})( {1 + {\breve{s}_i}}),\label{dec2b2}
\end{align}\label{dec2}
\end{subequations}\!\!\!based on which, the inference procedure is described by Algorithm 1. The associated analysis are presented in the following theorem, whose proof appears in Appendix F.

\begin{thm}
Consider inference procedure in Algorithm 1.  If the inferred subconscious bias ${\breve{s}}_i \neq 0$ for $\forall i \in \mathbb{V}$, Algorithm~1 generates the inference solution \eqref{infso}. \label{th2}
\end{thm}

\begin{algorithm}\small
  \caption{Inference from Estimation}
  \KwIn{Matrices $\widehat{\mathcal{A}}_{\vartheta}$ \eqref{esm} and vectors $\widehat{\mathfrak{a}}_{\vartheta}$ \eqref{esmok}, $\vartheta \!\in\! \{-1,+1\}$.}
    Subconscious bias:  ${{\breve{s}}_i} \!\leftarrow\! \frac{[\widehat{\mathfrak{a}}_{+1}]_{i} \!+\! [\widehat{\mathfrak{a}}_{-1}]_{i} \!+\! \sum\limits_{j \in \mathbb{V}} \!\!{( {{{[ \widehat{\mathcal{A}}_{+1}]}_{i,j}} - {{[ \widehat{\mathcal{A}}_{-1}]}_{i,j}}} )}}{2 - ([\widehat{\mathfrak{a}}_{+1}]_{i} \!- [\widehat{\mathfrak{a}}_{-1}]_{i} ) -\!\!  \sum\limits_{j \in \mathbb{V}}\! {( {{{[ \widehat{\mathcal{A}}_{+1}]}_{i,j}} \!+ {{[ \widehat{\mathcal{A}}_{-1}]}_{i,j}}} )}}$\;
   Confirmation bias: \!${{\breve{\epsilon}}_i} \!\leftarrow\! \frac{[\widehat{\mathfrak{a}}_{+1}]_{i} - [\widehat{\mathfrak{a}}_{-1}]_{i}}{4} \!-\! \frac{\sum\limits_{j \in \mathbb{V}} \!\!{( {{{[ \widehat{\mathcal{A}}_{+1}]}_{i,j}} - {{[ \widehat{\mathcal{A}}_{-1}]}_{i,j}}} )}}{4\breve{s}_{i}}$;\\
   Novelty bias: \!${{\breve{\eta}}_i} \!\leftarrow\! \frac{[\widehat{\mathfrak{a}}_{+1}]_{i} - [\widehat{\mathfrak{a}}_{-1}]_{i}}{4} \!+\! \frac{\sum\limits_{j \in \mathbb{V}}\!\! {( {{{[ \widehat{\mathcal{A}}_{+1}]}_{i,j}} - {{[ \widehat{\mathcal{A}}_{-1}]}_{i,j}}} )}}{4\breve{s}_{i}}$;\\
    Influence sum: $\sum\limits_{j \in \mathbb{V}} {{\breve{w}_{ij}}} \!\leftarrow\! \frac{\sum\limits_{j \in \mathbb{V}} {( {{{[ \widehat{\mathcal{A}}_{+1}]}_{i,j}} + {{[ \widehat{\mathcal{A}}_{-1}]}_{i,j}}} )}}{2} + {\breve{\eta}_i} - {\breve{\epsilon}_i}$;\\
   Topology: $\breve{w}_{ij} \!\leftarrow\! \begin{cases}
\!\frac{{[\widehat{\mathcal{A}}_{+1}]_{i,i}} + {[\widehat{\mathcal{A}}_{-1}]_{i,i}} -  2{\breve{\epsilon}_i}}{2(1 - \frac{{{\breve{\eta}_i}}}{{\sum\limits_{l \in \mathbb{V}} {{{{{\breve{w}_{il}}}}}} }})}  &\text{if}~i \!=\! j \\
\!  \frac{{[\widehat{\mathcal{A}}_{+1}]_{i,j}} + {[\widehat{\mathcal{A}}_{-1}]_{i,j}} }{2(1 - \frac{{{\breve{\eta}_i}}}{{\sum\limits_{l \in \mathbb{V}} {{{{{\breve{w}_{il}}}}}} }})} ,&\text{if}~i \!\neq\! j.
\end{cases}$
\end{algorithm}

\vspace{-0.25cm}
\section{Empirical Validation}
\vspace{-0.10cm}
In this section, we use US Senate Member Ideology data \cite{lewis2018voteview} to validate the theoretical results and model from perspectives of generalization error and model error. Since a senate member usually retires after (at most) twelve congresses, it is not practical to model the state member as an individual in our proposed opinion evolution model. Alternatively, an individual in our model represents one US state, and her opinion corresponds to the average of ideological data of senate members from the same state. Meanwhile, we model US President as information source in our model. To perform validation, we use the first-dimension ideological data obtained via Nokken-Poole estimation, which describes the economic liberalism-conservatism of a member. We consider the data of the 37th Congress to the 116th Congress, during which US President is from Republican Party or Democratic Party\footnotemarkmath\footnotetextmath{\textcolor[rgb]{0.00,0.00,1.00}{Presidents of the United States: \url{https://en.wikipedia.org/wiki/List_of_presidents_of_the_United_States}}}. However, the ideology of US President is not estimated in \cite{lewis2018voteview}. As an alternate, we set the default ideology of the president as $+1$ if the president is from the Republican Party, and $-1$ if the president is from the Democratic Party.

We make the worst-case assumptions on noise, i.e., $\sigma_{\mathrm{o}} = \sigma_{\mathrm{p}} = 1$. To guarantee Assumptions \ref{asprv}-3) and \ref{asprv}-4) hold, we let $\gamma = 1.305$, $\varrho_{1} = 0.12$ and $\varrho_{2} = 0.0001$. Since real system matrix is Schur stable, we can let ${\overline{\mathfrak{s}}} = 0.9$ and ${\underline{\mathfrak{s}}} = 0.1$. We set other parameters as $\varepsilon = 0.06$, $\rho = \frac{\varrho_{1}}{1.02}$, $\mathfrak{c} = 5.5$ and $\kappa = 2\sqrt{2}$.
\begin{figure}
\centering
\includegraphics[scale=0.34]{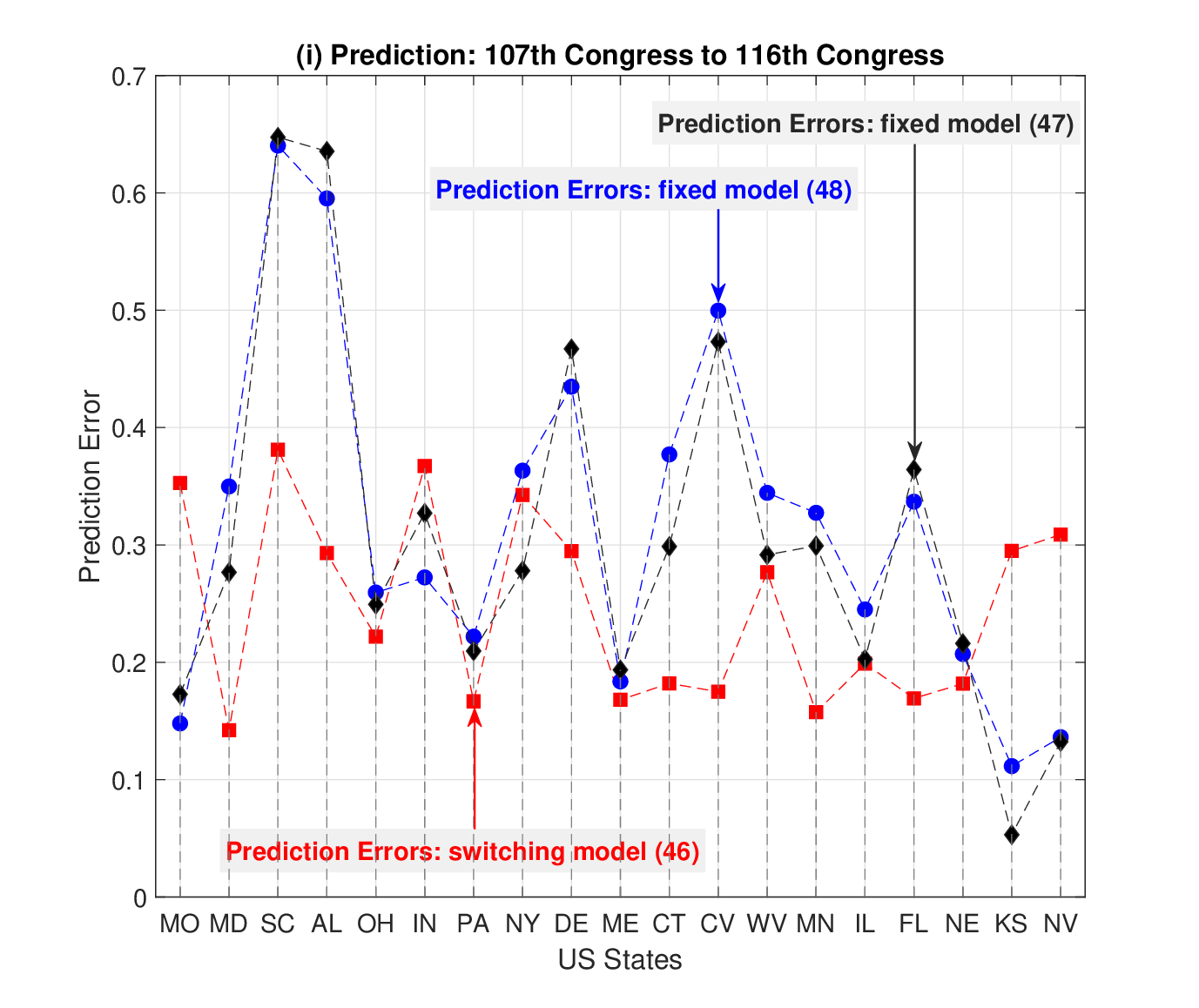}\\
\includegraphics[scale=0.345]{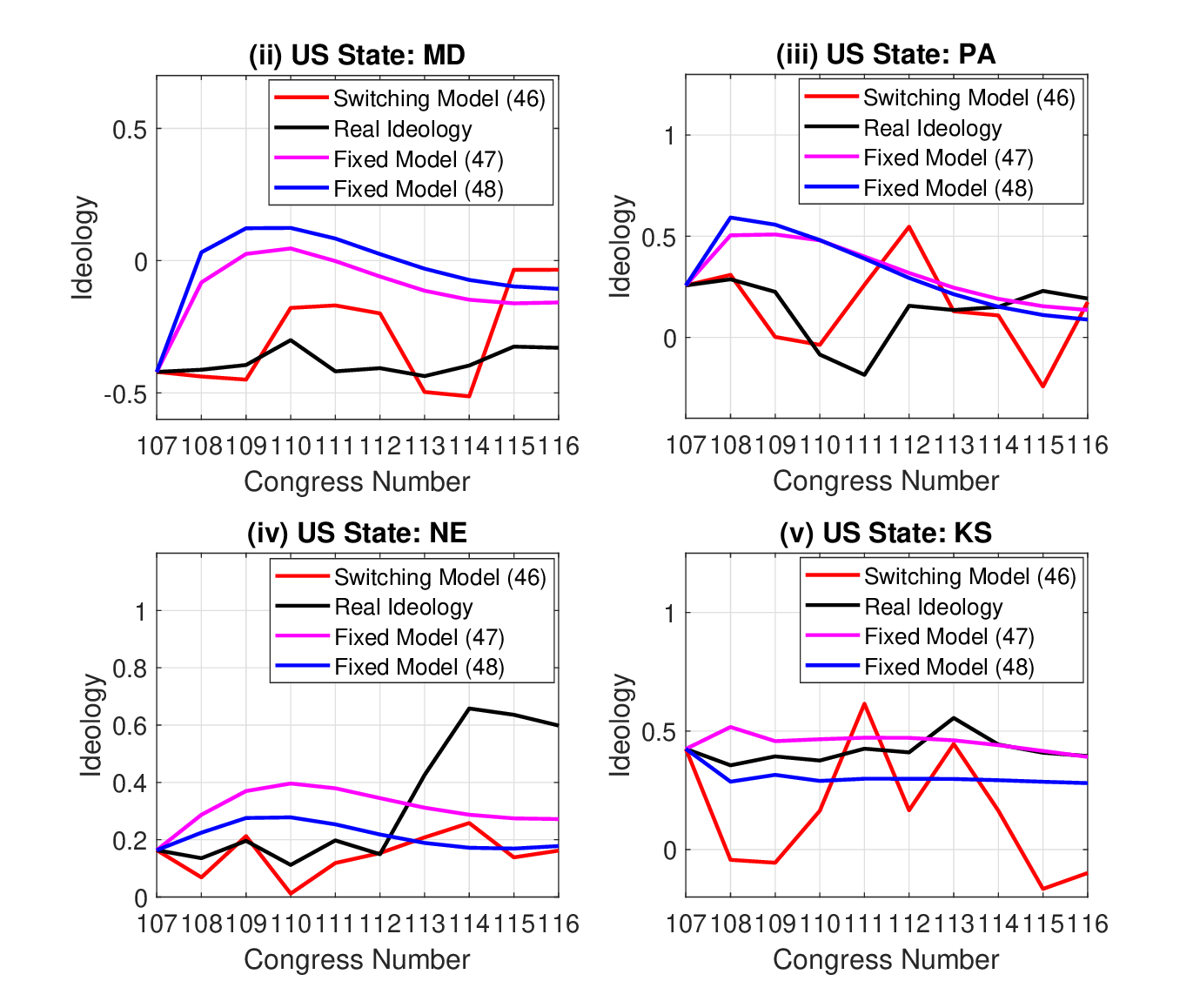}
\caption{(i): Nineteen states' model-based prediction errors; (ii)-(v): four states' real and predicted trajectories of ideology.}
\label{sp221}
\end{figure}

\vspace{-0.9cm}
\subsection{Prediction Error}
\vspace{-0.10cm}
The dynamics \eqref{sdw6} and \eqref{sdw4} show that when the information sources hold extremal opinions, the proposed model \eqref{kka} transforms to linear stochastic systems but have distinctive differences. This observation indicates that if the inferred social systems are leveraged for prediction, the inferred switching model that corresponds to \eqref{sdw6} and \eqref{sdw4}, i.e.,
\begin{subequations}
\begin{align}
y( {k + 1}) &= \widehat{\underline{\alpha}}  + \widehat{\underline{\mathcal{W}}}y(k),   ~~~\text{Democratic Party} \label{swp2}\\
y( {k + 1}) &= \widehat{\overline{\alpha}}  + \widehat{\overline{\mathcal{W}}}y(k),   ~~~\text{Republican Party} \label{swp1}
\end{align}\label{swpr}
\end{subequations}
\!\!would have smaller prediction error than that of the fixed social models, e.g.,
\vspace{-0.0cm}
\begin{align}
y( {k + 1}) &= \widehat{{\widetilde{\mathcal{W}}}}y(k),  ~~~~~~\text{DeGroot model \cite{degroot1974reaching}} \label{fmp0}\\
y( {k + 1}) &= \widehat{{\alpha}}  + \widehat{{\mathcal{W}}}y(k), ~\text{Friedkin-Johnsen model \cite{friedkin1990social}} \label{fmp}
\end{align}
which do not take confirmation bias and novelty bias into account.

We use the ideological data of 41st Congress to 106th Congress to infer models and save the rest of data (i.e., 107th--116th) to measure prediction error. Meanwhile, we assume we know that in the 107th--110th, 115th and 116th Congresses, US Presidents are from Republican Party, while in the 111th-114th Congresses, US Presidents are from Democratic Party. We follow the following procedure to perform the prediction.
\begin{itemize}
  \item We consider three group data: Republican Data (extracted if president is from Republican Party),  Democratic Data (extracted if president is from Democratic Party) and Mixed Data (no separation).
  \item We use the Democratic  Data, Republican Data and Mixed Data to respectively infer the sub-models \eqref{swp1} and \eqref{swp2} and the fixed models \eqref{fmp0} and \eqref{fmp}.
  \item For the prediction, we input the ideological data of the 107th Congress as the same initial condition for the switching model \eqref{swpr} and the models \eqref{fmp0} and \eqref{fmp}.
  \item From the 107th to 110th Congresses, we use model \eqref{swp1} for prediction, from the 111th to 114th Congresses, we switch to model \eqref{swp2} for prediction, in the 115th and 116th Congresses, we switch back to model \eqref{swp1}.
\end{itemize}
We note the sizes of Republican Data and Democratic Data are 38 and 28. \textcolor[rgb]{0.00,0.00,1.00}{Following $\underline{p}$, $\overline{k}$ and $\overline{p}$ given in \eqref{mmck}, we have
$\underline{p} = 28$, $\overline{k} = \underline{p} + 1 = 29$ and $\overline{p} = \overline{k} + 38 - 1 = 66$}. Then, following Corollary \ref{fth}, for the $(1.39,0.1)$-PAC, the allowed maximum network size is 19. Hence, we consider a network with 19 US states. We denote individual $\mathrm{v}_{i}$'s predicted ideology at congress number $k$ by $\breve{y}_{i}(k)$. We define the following metric to measure prediction error:
\begin{align}
{e_i} = \frac{1}{{10}}\sum\limits_{k = 107}^{116} {\left| {{y_i}(k) - {{\breve{y}}_i}(k)} \right|},  ~i \!\in\! \{\text{MO},\text{MD}, \ldots, \text{KS},\text{NV}\}. \nonumber
\end{align}
The nineteen states' prediction errors, and the picked four states' real ideology and predicted trajectories are respectively shown in Figure \ref{sp221} (i)-(v), observing which we discover that switching social model \eqref{swpr} has more accurate prediction than the fixed models \eqref{fmp0} and \eqref{fmp}, with exception being only total five states: MO, IN, NY, KS and NV.

\vspace{-0.305cm}
\subsection{Model Error and Fitting Error}
\vspace{-0.10cm}
Differentiating from numerical examples and man-made systems, we do not have real exact model parameters as references  to straightforwardly measure model error pertaining to $(\phi, \delta)$--PAC. Observing the matrices and vectors in \eqref{sdw7} and \eqref{sdw5} and recalling the convex combination \eqref{sdw3}, we can perform model validation from the following social system properties:
\begin{itemize}
  \item $0 \le {\overline{\alpha}_i} + \sum\limits_{j \in \mathbb{V}} {{{\left[ \overline{\mathcal{W}} \right]}_{i,j}}}  \le 1$,~~~$\forall i \in \mathbb{V}$.
  \item The magnitudes of all entries of $\overline{\mathcal{W}}$ and $\underline{\mathcal{W}}$ are smaller than one.
  \item The fitting curve and the trajectories of inferred model under arbitrary initial condition in $[-1,1]$, are all constrained into $[-1,1]$ for any time.
\end{itemize}

\begin{figure}
\centering
\includegraphics[scale=0.40]{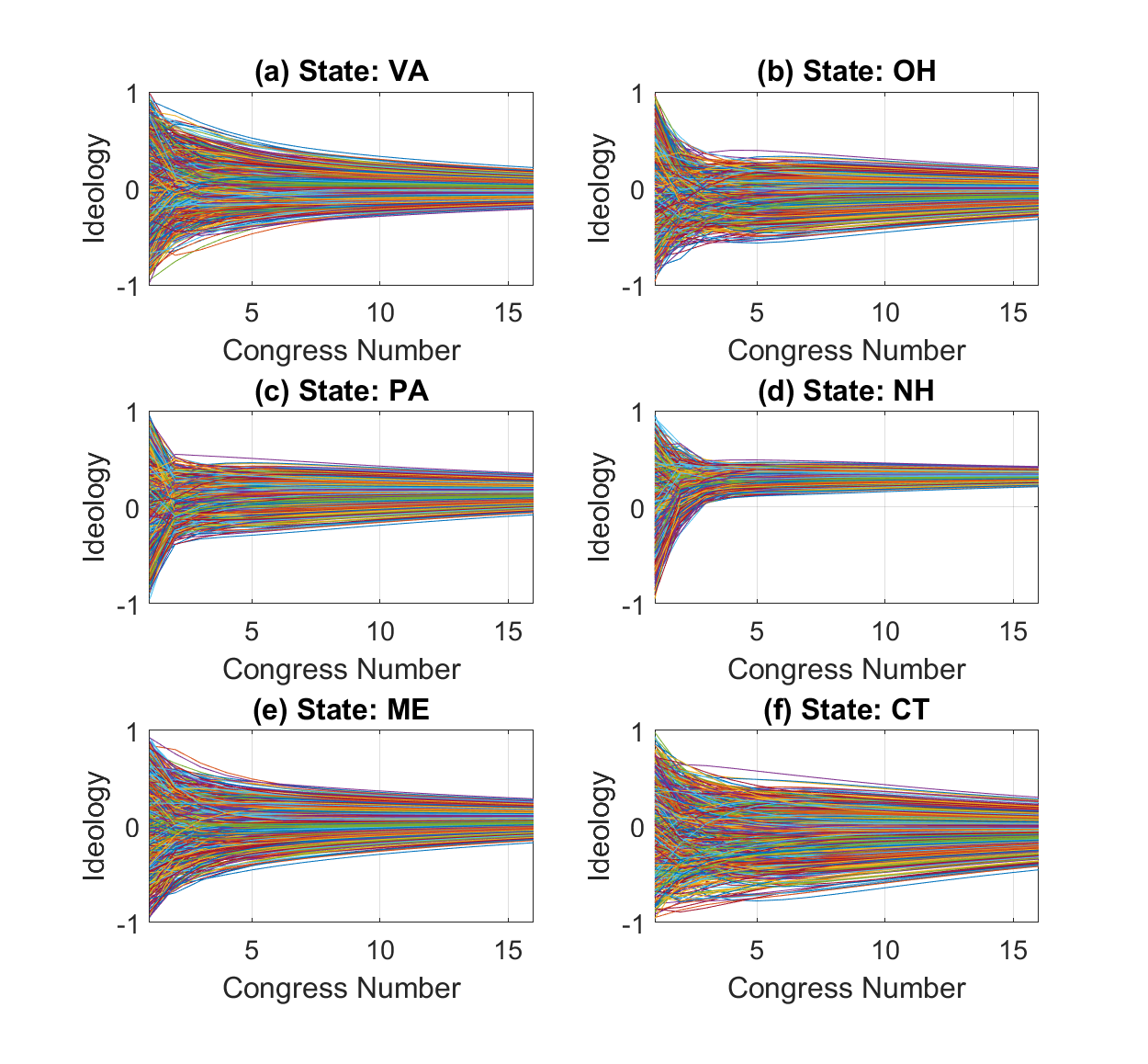}\\
\includegraphics[scale=0.36]{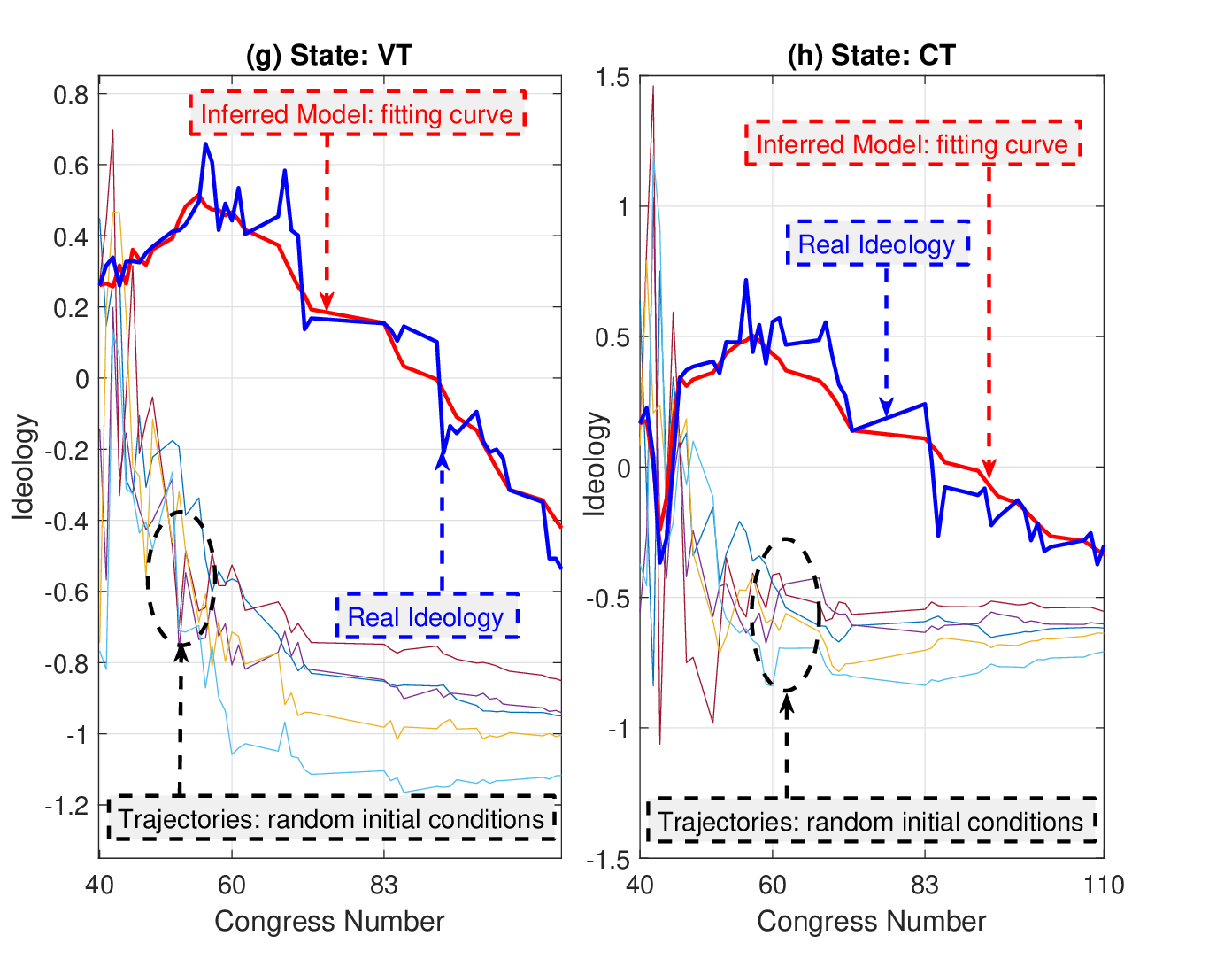}\\
\caption{(a)-(f): Evolving ideology under 1000 random initial conditions; (g)-(h): fitting curve and trajectories under five random initial conditions.}
\label{spcc}
\end{figure}

We now consider the ideological data of the 41st to 116th Congresses. The sizes of Republican  Data  and  Democratic  Data are 44 and 32, \textcolor[rgb]{0.00,0.00,1.00}{which means $\underline{p} = 32$ and $\overline{k} = \underline{p} + 1 = 33$ and $\overline{p} = \overline{k} + 44 - 1 = 76$}. Following Corollary \ref{fth}, for the $(0.27,0.1)$-PAC, the allowed network size is 6. Then, by \eqref{esm} and \eqref{esmok}, we have ${\widehat{\overline{\alpha}}}$ $=$ $[0.0156$, $-0.0461$, $0.0985$, $0.1476$, $0.0012$, $-0.0679]^{\top}$ and
\begin{align}\small
\widehat{\overline{\mathcal{W}}} \!=\!\! \left[\!\!\! {\begin{array}{*{21}{c}}
0.5252  \!\!&\!\!  0.0873  \!\!&\!\! -0.1641  \!\!&\!\!  0.0730 \!\!&\!\!  -0.3392  \!\!&\!\! -0.0049\\
-0.2307 \!\!&\!\!   0.4495 \!\!&\!\!  -0.0037 \!\!&\!\!   0.1190 \!\!&\!\!   0.0151  \!\!&\!\!  0.1655\\
-0.2311 \!\!&\!\!   0.0762  \!\!&\!\!  0.2090 \!\!&\!\!   0.1117 \!\!&\!\!   0.0155 \!\!&\!\!   0.1779\\
 0.0914 \!\!&\!\!   0.0030 \!\!&\!\!   0.1943 \!\!&\!\!   0.4089  \!\!&\!\!  0.0589  \!\!&\!\!  0.0940 \\
-0.1274 \!\!&\!\!  -0.0977 \!\!&\!\!   0.1578 \!\!&\!\!   0.0076 \!\!&\!\!   0.6119 \!\!&\!\!   0.1121\\
-0.2357 \!\!&\!\!   0.0309 \!\!&\!\!   0.1071 \!\!&\!\!   0.1157 \!\!&\!\!  -0.0139 \!\!&\!\!   0.7037
\end{array}} \!\!\!\right]\!\!,\nonumber
\end{align}
from which we verify that $|{[\widehat{\overline{\mathcal{W}}}]}_{i,j}| < 1$, $\forall i,j \in \mathbb{V}$, and
$\mathbf{0} < \widehat{\overline{\alpha}}$ + $\sum\limits_{j \in \mathbb{V}}{{[ \widehat{\overline{\mathcal{W}}}]}_{:,j}}$ =
$[0.1928$, $0.4685$, $0.4578$, $0.9982$, $0.6654$, $0.6399]^{\top} < \mathbf{1}$. Thus, the properties of social system matrix are demonstrated to hold. The trajectories of inferred model under 1000 randomly generated initial conditions in $[-1,1]$ are shown in Figure \ref{spcc} (a)-(f), which shows that all of the trajectories are constrained into $[-1,1]$. By Algorithm~1, some individuals are inferred to have novelty bias. The results together also demonstrate the correctness of the statement in Remark \ref{cococo}.

We next increase the size of social network to include all of the 30 states in the 40th Congress. We recall that the size of  Democratic Data is 32, which implies the maximum size of social network is 30. We thus can conclude the inference of the social network with 30 individuals hardly achieves any $\left(\phi, \delta\right)$--PAC. In this setting, the fitting curve and the five trajectories under random initial conditions in $[-1,1]$ are shown in Figure \ref{spcc} (g) and (h), which show that although the inferred model fit the real data well, without satisfying high PAC, the inferred model has larger model error such that its evolving ideologies under some initial conditions exceed the range $[-1,1]$ and the inferred model can be unstable.

\vspace{-0.12cm}
\section{Conclusion}
\vspace{-0.1cm}
In this paper, we have proposed an opinion evolution model which explicitly takes confirmation bias, novelty bias and process noise into account. Based on the proposed model,
we have studied the problem of social system inference of network topology, subconscious and model parameters of confirmation and novelty bias. We have analyzed the sample complexity of the proposed inference procedure in the presence of observation noise, which leads to the statistical dependence of observed public evolving opinions on time and coordinates. Real data validations suggest the effectiveness of the obtained theoretical results and the proposed opinion evolution model.

In the future research, we will investigate the sample complexity of social system inference in the scenario that process and observation noise have time-varying means and variances.

\vspace{-0.25cm}
\appendices
\vspace{-0.05cm}
\section*{Appendix A: Notations}
\vspace{-0.20cm}
The auxiliary vector notations are defined by:
\vspace{-0.2cm}
\begin{subequations}
\begin{align}
\hspace{-0.81cm}{{\mathbf{o}}_{\vartheta}} &\triangleq [ {\mathbf{o}_{k_{\vartheta}};~ \mathbf{o}_{k_{\vartheta}+1};~ \ldots;~ {\mathbf{o}}_{p_{\vartheta} - 3};~ {\mathbf{o}}_{p_{\vartheta} - 2}}],\label{ghb3}\\
\hspace{-0.81cm}{{\mathbf{o}}_{k_{\vartheta}}} &\triangleq  [ {\widetilde{\mathfrak{o}}^{k_{\vartheta}+1}_{k_{\vartheta}};~\widetilde{\mathfrak{o}}^{k_{\vartheta}+2}_{k_{\vartheta}};~\ldots ;~\widetilde{\mathfrak{o}}^{p_{\vartheta}-1}_{k_{\vartheta}}}],\label{ghb2}
\end{align}\label{asv1}
\end{subequations}
\vspace{-0.85cm}
\begin{subequations}
\begin{align}
{{\mathbf{p}}_{\vartheta}} &\triangleq [ {{{{\mathbf{p}}}_{k_{\vartheta}}};~ {{{\mathbf{p}}}_{k_{\vartheta}+1}};~ \ldots;~{{{\mathbf{p}}}_{p_{\vartheta}-3}};~{{{\mathbf{p}}}_{p_{\vartheta}-2}}}],\label{ghb6}\\
{{{\mathbf{p}}}_{i_{\vartheta}}} &\triangleq [ {{{\widetilde{\mathbf{p}}}_{i_{\vartheta}+1}};~{{\widetilde{\mathbf{p}}}_{i_{\vartheta}+2}};~\ldots;~{{\widetilde{\mathbf{p}}}_{p_{\vartheta}-2}};~{{\widetilde{\mathbf{p}}}_{p_{\vartheta}-1}}}],\label{ghb5}\\
{{\widetilde{\mathbf{p}}}_{j}} &\triangleq [ {\mathfrak{p}(1 ),~\ldots,~\mathfrak{p}( j\!-\!2 ),~\mathfrak{p}( j\!-\!1)}],\label{ghb4}
\end{align}\label{asv2}
\end{subequations}
\vspace{-0.90cm}
\begin{subequations}
\begin{align}
\hspace{-0.10cm}{{\mathbf{a}}_{\vartheta}} &\triangleq [ {{{{\mathbf{a}}}_{k_{\vartheta}}};~ {{{\mathbf{a}}}_{k_{\vartheta}+1}};~ \ldots;~{{{\mathbf{a}}}_{p_{\vartheta}-3}};~{{{\mathbf{a}}}_{p_{\vartheta}-2}}}],\label{ghba6}\\
\hspace{-0.10cm}{{{\mathbf{a}}}_{i_{\vartheta}}} &\triangleq [ {{{\widetilde{\mathbf{a}}}_{i_{\vartheta}+1}};~{{\widetilde{\mathbf{a}}}_{i_{\vartheta}+2}};~\ldots;~{{\widetilde{\mathbf{a}}}_{p_{\vartheta}-2}};~{{\widetilde{\mathbf{a}}}_{p_{\vartheta}-1}}}],\label{ghba5}\\
\hspace{-0.10cm}{{\widetilde{\mathbf{a}}}_{j}} &\triangleq [ {\alpha_{\vartheta(1)};~\ldots;~\alpha_{\vartheta( j-2 )};~\alpha_{\vartheta( j-1 )}}],\label{ghba4}
\end{align}\label{asv3}
\end{subequations}
\vspace{-0.90cm}
\begin{align}
\hspace{-0.40cm}\mathbf{x}_{\vartheta} &\triangleq \left[ {{x}(1);{x}(1);{x}(1);{x}(1);\ldots; {x}(1)} \right] \in {\mathbb{R}^{{\mathfrak{l}_{\vartheta}}}}.\label{asv4}
\end{align}
\vspace{-0.7cm}

The auxiliary matrix notations are defined by:
\vspace{-0.2cm}
\begin{align}
{\breve{\overline{\mathrm{W}}}}_{\vartheta} \triangleq  {\text{diag}\{{\mathcal{M}_{g}^\mathrm{c} - \mathcal{M}_{j}^\mathrm{c}}\}_{g<j\in\{k_{\vartheta},\ldots,p_{\vartheta}-1\}}} \in \mathbb{R}^{\mathfrak{l}_{{\vartheta}} \times \mathfrak{l}_{{\vartheta}}}, \label{gh2}
\end{align}
\vspace{-0.80cm}
\begin{subequations}
\begin{align}
\!\!\!\!\!&{\overline{\mathrm{W}}}_{\vartheta} \!\triangleq\!   {\text{diag}\{ {{\overline{\mathrm{A}}_{k_{\vartheta}}},~{\overline{\mathrm{A}}_{k_{\vartheta}+1}},~\ldots,~{\overline{\mathrm{A}}_{p_{\vartheta}-3}},~{\overline{\mathrm{A}}_{p_{\vartheta}-2}}}\}}, \label{gh5}\\
\!\!\!\!\!&{\overline{\mathrm{A}}_{k_{\vartheta}}} \!\triangleq\!  {\text{diag}\{ {{\overline{\mathrm{A}}_{(k_{\vartheta},k_{\vartheta}+1)}},{\overline{\mathrm{A}}_{(k_{\vartheta},k_{\vartheta}+2)}},\ldots,{\overline{\mathrm{A}}_{(k_{\vartheta},p_{\vartheta}-1)}}} \}}, \label{gh4}\\
\!\!\!\!\!&{\overline{\mathrm{A}}_{(g,j)}} \!\triangleq\! [{\mathcal{M}_{( {g - 1,g})}}(\mathbf{I}_{n} - \mathcal{M}_{(j - g+1,j)}), ~\ldots, \nonumber\\
&\hspace{1.42cm}{\mathcal{M}_{({1,g})}}(\mathbf{I}_{n} - \mathcal{M}_{(j - g+1,j)}),~-{\mathcal{M}_{( {j - g,j})}},\nonumber\\
&\hspace{1.42cm}-{\mathcal{M}_{( {j - g - 1,j})}},~\ldots,~-{\mathcal{M}_{( {1,j})}}], ~~~g < j.\label{gh3}
\end{align}\label{ghmm2}
\end{subequations}

\vspace{-0.85cm}
\section*{Appendix B: Auxiliary Lemmas}
\vspace{-0.15cm}
\begin{lem}\cite{adamczak2015note}
Let $\mathbf{f}$ be a mean zero random vector in $\mathbb{R}^{n}$, whose covariance matrix is denoted by $\mathrm{Cov}( \mathbf{f} )$. If $\mathbf{f}$ has the convex concentration property with constant $\kappa$, then for any $A \in \mathbb{R}^{n \times n}$ and every $t > 0$, we have
\vspace{-0.3cm}
\begin{align}
\mathbf{P}\!\!\left[ \left| {{\mathbf{f}^\top}A\mathbf{f} - \mathbf{E}[ {{\mathbf{f}^\top}A\mathbf{f}}]} \right| \ge t\right] &\le 2{e^{{ \textcolor[rgb]{0.00,0.00,1.00}{- \frac{1}{{\mathfrak{c}{\kappa^2}}}\min \left\{ {\frac{{{t^2}}}{{\left\| A \right\|_{\mathrm{F}}^2\mathrm{Cov}( \mathbf{f} )}},~\frac{t}{{\left\| A \right\|}}}\right\}}}}}\! \nonumber
\end{align}
\vspace{-0.00cm}
for some universal constant $\mathfrak{c}$.  \label{newr}
\end{lem}
\begin{lem} \textcolor[rgb]{0.00,0.00,1.00}{[Chapter 4, \cite{vershynin2018high}]} Let $W$ be an $d \times d$ a symmetric random matrix. Furthermore, let $\mathcal{N}$ be an $\varepsilon$-net of $\mathcal{S}^{d-1}$ with minimal cardinality. Then for all $\rho > 0$, we have
\vspace{-0.25cm}
\begin{align}
&\mathbf{P}\left[|| W || > \rho \right] \nonumber\\
&\le {( {\frac{2}{\varepsilon } + 1})^{\textcolor[rgb]{0.00,0.00,1.00}{d}}}\mathop {\max }\limits_{\mathbf{u} \in \mathcal{N}} \mathbf{P}\left[ {|| W\mathbf{u} ||_{2} > ( {1 - \varepsilon })\rho }\right], ~\varepsilon \in [0, 1).\label{aalem1}\\
&\mathbf{P}\left[|| W || > \rho \right] \nonumber\\
&\le {( {\frac{2}{\varepsilon } \!+\! 1})^{\textcolor[rgb]{0.00,0.00,1.00}{d}}}\mathop {\max }\limits_{\mathbf{u} \in \mathcal{N}} \mathbf{P}\left[ {\left| {{\mathbf{u}^\top}W\mathbf{u}} \right| \!>\! ( {1 \!-\! 2\varepsilon })\rho }\right]\!, ~\varepsilon \in [0, \frac{1}{2}). \label{aalem2}
\end{align}\label{cko}
\end{lem}
\vspace{-0.50cm}
\begin{lem}~\cite{abbasi2011improved}
Let $\{\mathcal{F}_{t}\}_{t \geq 1}$ be a filtration. Let $\{\eta_{t}\}_{t \geq 1}$ be a stochastic process adapted to $\{\mathcal{F}_{t}\}_{t \geq 1}$ and taking values in $\mathbb{R}$. Let $\{x_{t}\}_{t \geq 1}$ be a predictable stochastic process with respect to $\{\mathcal{F}_{t}\}_{t \geq 1}$, taking values in $\mathbb{R}^{d}$. Furthermore, assume that $\eta_{t}$ is $\mathcal{F}_{\textcolor[rgb]{0.00,0.00,1.00}{t}}$-measurable and conditionally $\gamma$-sub-Gaussian for some $\gamma > 0$. Let $S > 0$, $\eta^{\top} = [\eta_{2},\eta_{3}, \ldots,\eta_{t+1}]$, and $X^{\top} = [x_{1},x_{2}, \ldots,x_{t}]$. The following
\vspace{-0.25cm}
\begin{align}
||{{{\left( {{X^ \top }X \!+\! S} \right)}^{ - 0.5}}{X^ \top }\eta } ||_2^2 \le 2{\gamma ^2}\!\ln\!{\frac{{{{\left( {\det\! \left({\left( {{X^ \top }X \!+\! S} \right)\!{S^{ - 1}}} \right)}\right)}^{0.5}}}}{\delta }} \nonumber
\end{align}
holds with the probability of at least $1-\delta$. \label{ckpq}
\end{lem}

\section*{Appendix C: Proof of Proposition \ref{ptopk1a}}
\vspace{-0.1cm}
It follows from \eqref{oEp2} that
\begin{align}
\mathbf{E}[{\Psi_{\vartheta}^\top {X_{\vartheta}}X_{\vartheta}^\top {\Psi_{\vartheta}}}] = \mathbf{I}_{n}, \label{hhab}
\end{align}
by which, we obtain that
\begin{align}
&\left\| \varrho_{2}{{{( {X_{\vartheta}^ \top {\Psi_{\vartheta}}})^\top}}X_{\vartheta}^ \top {\Psi_{\vartheta}} - \varrho_{1}\mathbf{I}_{n}} \right\|\nonumber\\
& \!=\! \left\| {{{\varrho_{2}( {X_{\vartheta}^ \top {\Psi_{\vartheta}}})^\top} }X_{\vartheta}^ \top {\Psi_{\vartheta}}} \right. \left. { - \varrho_{1}\mathbf{E}[ {\Psi_{\vartheta}^ \top {X_{\vartheta}}X_{\vartheta}^ \top {\Psi_{\vartheta}}}]} \right\| \nonumber\\
& \!=\!\! \mathop {\sup }\limits_{\mathbf{u} \in {\mathcal{S}^{n - 1}}} \!\!\left| \varrho_{2}{{\mathbf{u}^\top}{{( {X_{\vartheta}^\top {\Psi_{\vartheta}}})^\top}}X_{\vartheta}^ \top {\Psi_{\vartheta}}\mathbf{u}} \right. \left. { \!- \varrho_{1}{\mathbf{u}^\top}\mathbf{E}[ {\Psi_{\vartheta}^\top {X_{\vartheta}}X_{\vartheta}^\top {\Psi_{\vartheta}}}]\mathbf{u}} \right|\nonumber\\
& \!=\!\! \mathop {\sup }\limits_{\mathbf{u} \in {\mathcal{S}^{n - 1}}} \!\!\left| {\varrho_{2}|| {X_{\vartheta}^ \top\! {\Psi_{\vartheta}}\mathbf{u}} ||_2^2 - \varrho_{1}\mathbf{E}|| {X_{\vartheta}^ \top \!{\Psi_{\vartheta}}\mathbf{u}} ||_2^2} \right|\!.   \label{pqc}
\end{align}

We obtain from \eqref{traa}--\eqref{dp1} that
\vspace{-0.1cm}
\begin{align}
\widetilde{\mathbf{y}}^{j}_{g} &\!=\! ( {\mathcal{M}_{g}^\mathrm{c} \!-\! \mathcal{M}_{j}^\mathrm{c}})x(1) \!-\! \sum\limits_{i = 1}^{j - g}\!{\mathcal{M}_{(i,j)}}( {{\alpha _{\vartheta(j-i)}} \!+\! \mathfrak{p}(j\!-\!i)})  \!+\! \mathfrak{o}_g^j \nonumber\\
&\hspace{0.41cm}\!\! \!+\! \sum\limits_{i = 1}^{g - 1}\! {\mathcal{M}_{(i,g)}}(\mathbf{I}_{n} \!-\! \mathcal{M}_{(j - g+1,j)}\!)( {{\alpha _{\vartheta(g-i)}} \!+\! \mathfrak{p}(g\!-\!i)}).\label{tr1}
\end{align}
Then, observing $\mathbf{U}$ in \eqref{adef1kk}, ${{\Upsilon}_{\vartheta}}$ in \eqref{adef1}, $X_{\vartheta}$ in \eqref{xf211} and the relation \eqref{ghhbbc},
we have
\begin{align}
X_{\vartheta}^\top {\Psi_{\vartheta}}\mathbf{u} = \mathbf{U}^{\top}{{\Upsilon}^{\top}_{\vartheta}}{{\Pi}_{\vartheta}}({{\overline{\theta}}_{\vartheta}} + {{\underline{\theta}}_{\vartheta}}), \label{ppmo}
\end{align}
substituting which into \eqref{pqc}, we arrive at
\begin{align}
&|| \varrho_{2}{{{( {X_{\vartheta}^ \top {\Psi_{\vartheta}}})^\top} }X_{\vartheta}^ \top {\Psi_{\vartheta}} - \varrho_{1}\mathbf{I}_{n}} ||  \nonumber\\
& = \mathop {\sup }\limits_{\mathbf{u} \in {\mathcal{S}^{n - 1}}} \left| {\varrho_{2}|| \mathbf{U}^{\top}{{\Upsilon}^{\top}_{\vartheta}}{{\Pi}_{\vartheta}}{({{\overline{\theta}}_{\vartheta}} + {{\underline{\theta}}_{\vartheta}})} ||_2^2} \right. \nonumber\\
&\hspace{3.2cm} \left. { - \varrho_{1}\mathbf{E}|| \mathbf{U}^{\top}{{\Upsilon}^{\top}_{\vartheta}}{{\Pi}_{\vartheta}}({{\overline{\theta}}_{\vartheta}} + {{\underline{\theta}}_{\vartheta}})||_2^2} \right|\!. \label{relatp}
\end{align}

We note that \eqref{hhab}, in conjunction with \eqref{pqc} and \eqref{ppmo}, implies $\mathbf{E}||{X_{\vartheta}^\top {\Psi_{\vartheta}}\mathbf{u}} ||_2^2 = \mathbf{E}|| \mathbf{U}^{\top}{{\Upsilon}^{\top}_{\vartheta}}{{\Pi}_{\vartheta}}({{\overline{\theta}}_{\vartheta}} + {{\underline{\theta}}_{\vartheta}})||_2^2 = 1$, which thus with \eqref{relatp} and Assumption \ref{asprv}-4) indicate that
\begin{align}
&\left| {||{{\bf{U}}^ \top }\Upsilon _{\vartheta}^ \top {\Pi _{\vartheta}}{\overline{\theta}_{\vartheta}}||_2^2 - {\bf{E}}||{{\bf{U}}^ \top }\Upsilon _{\vartheta}^ \top {\Pi _{\vartheta}}{\overline{\theta} _{\vartheta}}||_2^2} \right| \nonumber\\
& = \left| {||{{\bf{U}}^ \top }\Upsilon _{\vartheta}^ \top {\Pi _{\vartheta}}{\overline{\theta}_{\vartheta}}||_2^2} \right. + ||{{\bf{U}}^ \top }\Upsilon _{\vartheta}^ \top {\Pi_{\vartheta}}{\underline{\theta} _{\vartheta}}||_2^2  \nonumber\\
&\hspace{3.0cm}\left. { - {\bf{E}}||{{\bf{U}}^ \top }\Upsilon _{\vartheta}^ \top {\Pi _{\vartheta}}({\overline{\theta}_{\vartheta}} + {\underline{\theta}_{\vartheta}})||_2^2} \right|\label{relatmc1}\\
& \ge \left| {\varrho_{2}||{{\bf{U}}^\top }\Upsilon _{\vartheta}^ \top {\Pi _{\vartheta}}({\overline{\theta }_{\vartheta}} + {\underline{\theta }_{\vartheta}})||_2^2} \right. \nonumber\\
&\hspace{3.0cm}\left. { - \varrho_{1}{\bf{E}}||{{\bf{U}}^ \top }\Upsilon _{\vartheta}^ \top {\Pi _{\vartheta}}({\overline{\theta }_{\vartheta}} + {\underline{\theta }_{\vartheta}})||_2^2} \right|, \label{relatmc2}
\end{align}
where \eqref{relatmc1} from previous step is obtained via considering
\begin{align}
{\bf{E}}||\!{{\bf{U}}^\top}\!\Upsilon _{\vartheta}^\top\! {\Pi _{\vartheta}}({\overline{\theta}_{\vartheta}} \!+\! {\underline{\theta}_{\vartheta}})||_2^2 \!=\! {\bf{E}}||\!{{\bf{U}}^ \top}\!\Upsilon _{\vartheta}^\top\! {\Pi _{\vartheta}}\overline{\theta}_{\vartheta}||_2^2 \!+\!\! ||\!{{\bf{U}}^\top}\!\Upsilon _{\vartheta}^\top \!{\Pi _{\vartheta}}{\underline{\theta}_{\vartheta}}||_2^2\nonumber
\end{align}
which is due to Assumption \ref{asprv}-1) and definitions in \eqref{tr1aas}.

Let us define:
\begin{align}
\Delta_{\vartheta}  \triangleq {{\Pi}^{\top}_{\vartheta}}{{\Upsilon}_{\vartheta}}\mathbf{U}\mathbf{U}^{\top}{{\Upsilon}^{\top}_{\vartheta}}{{\Pi}_{\vartheta}}, \label{ghb1okok}
\end{align}
by which, we obtain
\textcolor[rgb]{0.00,0.00,1.00}{\begin{align}
|| {{\Delta _{\vartheta}}} ||_{\mathrm{F}}^2 &   \leq {|| {\Pi _{\vartheta}^ \top {\Upsilon _{\vartheta}}{\mathbf{U}}} ||^2}|| {{{\mathbf{U}}^ \top }\Upsilon _{\vartheta}^ \top {\Pi _{\vartheta}}} ||_{\mathrm{F}}^2 \label{ckj1}\\
&  = {|| {\Pi _{\vartheta}^ \top {\Upsilon _{\vartheta}}{\mathbf{U}}} ||^2}|| {\Pi _{\vartheta}^ \top {\Upsilon _{\vartheta}}{\mathbf{U}}}||_{\mathrm{F}}^2 \label{ckj2}\\
& \leq {|| {\Pi _{\vartheta}^ \top\! {\Upsilon _{\vartheta}}}||^2}{|| {\mathbf{U}} ||^2}{||{\Pi _{\vartheta}^\top\! {\Upsilon _{\vartheta}}} ||^2}|| {\mathbf{U}} ||_{\mathrm{F}}^2 \label{ckj3}\\
&  = {|| {\Pi _{(k,p)}^ \top {\Upsilon _{\vartheta}}} ||^4}||\frac{\mathfrak{l}_{\vartheta}}{n},  \label{ckj4}\\
||\Delta_{\vartheta}|| &\leq ||{\Pi^{\top}_{\vartheta}}{{\Upsilon}_{\vartheta}}||^{2}|| {\mathbf{U}} ||^2 = ||{\Pi^{\top}_{\vartheta}}{{\Upsilon}_{\vartheta}}||^{2},\label{ckj5}
\end{align}
where \eqref{ckj1}--\eqref{ckj3} are obtained via considering the well-known inequalities $|| {AB} ||$ $\le$ $|| A|||| B ||$, $|| {AB} ||_{\mathrm{F}}$ $\le$ $|| A |||| B ||_{\mathrm{F}}$ and $|| A ||_{\mathrm{F}} = || A^{\top} ||_{\mathrm{F}}$; \eqref{ckj4}  is obtained from \eqref{ckj3} via considering $|| {\mathbf{U}} ||^2 = 1$ and $|| {\mathbf{U}} ||_{\mathrm{F}}^2 = \frac{\mathfrak{l}_{\vartheta}}{n}$ (which is implied by $\mathbf{u} \in \mathcal{S}^{n-1}$ and \eqref{adef1kk}); and \eqref{ckj5} follows from $|| {AB} ||$ $\le$ $|| A|||| B ||$, $|| A || = || A^{\top} ||$ and $|| {\mathbf{U}} || = 1$.} With $\rho > 0$, from \eqref{ckj4} and \eqref{ckj5} we have
\begin{align}
&\min \left\{ {\frac{{{\rho ^2}}}{{|| \Delta_{\vartheta} ||_\mathrm{F}^2|| {\overline{\mathcal{C}}}_{\vartheta} ||}},~\frac{{\rho}}{{|| \Delta_{\vartheta} ||}}} \right\} \nonumber\\
&\geq \min \left\{{\frac{{{n\rho^2}}}{{\mathfrak{l}_{\vartheta}||\Pi _{\vartheta}^ \top {\Upsilon _{\vartheta}}||^4||{{\overline{\mathcal{C}}}_{\vartheta}}||}}} \right., \left. {\frac{\rho }{{||\Pi_{\vartheta}^ \top{\Upsilon _{\vartheta}}||^2}}} \!\right\}.\label{ghb1ok}
\end{align}

Under Assumption \ref{asprv}-1), we first verify from \eqref{tr1aas} with \eqref{asv1}-\eqref{asv4} that ${\overline{\theta}_{\vartheta}}$ has zero mean. Since $\mathfrak{c} > 0$ and $\rho > 0$, under Assumption \ref{asprv}-2), applying Lemma \ref{newr} (in Appendix B) with \eqref{ghb1ok} and \eqref{ghb1okok}, we conclude that
\begin{align}
\left| {||{\bf{U}}^\top  {{\Upsilon}_{\vartheta}^\top {{\Pi}_{\vartheta}}{{\overline{\theta}}_{\vartheta}}}||_2^2 - {\bf{E}}||{\bf{U}}^\top  {{\Upsilon}_{\vartheta}^\top {{\Pi}_{\vartheta}}{{\overline{\theta}}_{\vartheta}}} ||_2^2} \right| > \rho \label{mcp1}
\end{align}
holds with probability at most
\begin{align}
2e^{\frac{-1}{{\mathfrak{c}{\kappa^2}}}\!\min \left\{\! {\frac{{{n\rho^2}}}{{\mathfrak{l}_{\vartheta}||\Pi _{\vartheta}^ \top \!\!{\Upsilon _{\vartheta}}\!||^4||{{\overline{\mathcal{C}}}_{\vartheta}}\!||}}} \right.\!, \left. {\frac{\rho }{{||\Pi_{\vartheta}^ \top\!\!{\Upsilon _{\vartheta}}\!||^2}}} \!\right\}}\!, \label{mcp2}
\end{align}
where ${\overline{\mathcal{C}}}_{\vartheta} \triangleq  {\bf{E}}\left[{\overline{\theta}}_{\vartheta}{\overline{\theta}}^{\top}_{\vartheta}\right]$. It follows from ${\overline{\theta}}_{\vartheta}$ in \eqref{tr1aas} and ${\mathcal{C}}_{\vartheta}$ in \eqref{kop2ab} that $||{\overline{\mathcal{C}}}_{\vartheta}|| = ||{{\mathcal{C}}}_{\vartheta}||$, noting which, \eqref{relatp}, \eqref{relatmc2} and \eqref{ghb1ok} we conclude that
\begin{align}
&|| {{\varrho_{2}( {X_{\vartheta}^ \top {\Psi_{\vartheta}}})^\top}X_{\vartheta}^ \top {\Psi_{\vartheta}} - \varrho_{1}{\bf{I}_{n}}} || \nonumber\\
& =\mathop {\sup }\limits_{\mathbf{u} \in {\mathcal{S}^{n - 1}}} \!\!| {{\mathbf{u}^\top}\!\!( {{{\varrho_{2}(\! {X_{\vartheta}^\top {\Psi_{\vartheta}}})^\top} }\!X_{\vartheta}^\top {\Psi_{\vartheta}} - \varrho_{2}{\bf{I}_{n}}})\mathbf{u}}| \ge (1 - 2\varepsilon )\rho \nonumber
\end{align}
holds with probability at most
\begin{align}
2e^{\frac{-1}{{\mathfrak{c}{\kappa^2}}}\!\min \left\{{\frac{{{n(1 - 2\varepsilon )^2\rho^2}}}{{\mathfrak{l}_{\vartheta}||\Pi _{\vartheta}^ \top {\Upsilon _{\vartheta}}||^4||{{{\mathcal{C}}}_{\vartheta}}||}}} \right., ~\left. {\frac{(1 - 2\varepsilon )\rho }{{||\Pi_{\vartheta}^ \top\!\!{\Upsilon _{\vartheta}}||^2}}} \!\right\}}, \nonumber
\end{align}
Then, applying \eqref{aalem2} in Lemma \ref{cko} leads to Proposition \ref{ptopk1a}.

\vspace{-0.1cm}
\section*{Appendix D: Proof of Theorem \ref{ptopk1er}}
\vspace{-0.1cm}
Observing the relation \eqref{datanew11} and the optimal estimation \eqref{esm}, we obtain $\widehat{\mathcal{A}}_{\vartheta} \!-\! {{\mathcal{A}}_{\vartheta}}$ $\!=\!$ ${{U}}_{\vartheta}{X}^{\top}_{\vartheta}{({X}_{\vartheta}{X}^{\top}_{\vartheta})^{ - 1}}$. Thus,
\begin{align}
|| \widehat{\mathcal{A}}_{\vartheta} - {{\mathcal{A}}_{\vartheta}} || & \!=\! || {{U}}_{\vartheta}{X}^{\top}_{\vartheta}{({X}_{\vartheta}{X}^{\top}_{\vartheta})^{ - 1}} || \nonumber\\
& \!\le\! || {{U}}_{\vartheta}{X}^{\top}_{\vartheta}{({X}_{\vartheta}{X}^{\top}_{\vartheta})^{ - 0.5}} |||| {{{( {{{X}_{\vartheta}}X_{\vartheta}^ \top })^{ - 0.5}}}} ||, \nonumber
\end{align}
for which we define two events:
\begin{align}
{\mathfrak{E}_1} &\triangleq \left\{ || {{U}}_{\vartheta}{X}^{\top}_{\vartheta}{({X}_{\vartheta}{X}^{\top}_{\vartheta})^{ - 0.5}} || \right.\left. { \!\!|| {{{( {{{X}_{\vartheta}}X_{\vartheta}^ \top })^{ - 0.5}}}} || > \phi } \right\},\label{event1}\\
{\mathfrak{E}_2} &\triangleq \left\{ {|| {\varrho_{2}{\Psi^\top_{\vartheta}} {X_{\vartheta}}X_{\vartheta}^ \top {{\Psi_{\vartheta}}} - \varrho_{1}\mathbf{I}_{n}} || \le \rho } \right\},\label{event2}
\end{align}
from which we have
\begin{align}
\mathbf{P}[|| \widehat{\mathcal{A}}_{\vartheta} - {{\mathcal{A}}_{\vartheta}} || > \phi ] \le \mathbf{P}[{\mathfrak{E}_1}\bigcap {{\mathfrak{E}_2}}] + \mathbf{P}[\mathfrak{E}_2^\mathrm{c}]. \label{kp1}
\end{align}
We next derive the upper bounds on $\mathbf{P}( {\mathfrak{E}_2^\mathrm{c}})$ and $\mathbf{P}( {{\mathfrak{E} _1}\bigcap {{\mathfrak{E} _2}}})$.

\subsubsection*{\underline{Upper Bound on $\mathbf{P}( {\mathfrak{E}_2^\mathrm{c}})$}} Let us set $\gamma  = \sqrt {2\mathfrak{c}} \kappa$, inserting which into \eqref{nc1} results in
\begin{align}
\min \left\{{\frac{{{{(1 - 2\varepsilon )}^2}n{\rho ^2}}}{{\mathfrak{l}_{\vartheta}||\Pi _{\vartheta}^\top {\Upsilon _{\vartheta}}||^4||{\mathcal{C}_{\vartheta}}||}},\frac{{(1 - 2\varepsilon )\rho }}{{||\Pi_{\vartheta}^ \top {\Upsilon _{\vartheta}}||^2}}} \!\right\} \ge \mathfrak{c}{\kappa ^2}\ln \frac{{4 \cdot {( {\frac{2}{\varepsilon } + 1})^n}}}{\delta }, \nonumber
\end{align}
which is equivalent to
\begin{align}
2 \cdot {( {\frac{2}{\varepsilon } + 1})^n} \cdot {e^{ \frac{- 1}{{\mathfrak{c}{\kappa ^2}}}\min \left\{ {\frac{{{{(1 - 2\varepsilon )}^2}n{\rho ^2}}}{{\mathfrak{l}_{\vartheta}||\Pi _{\vartheta}^\top {\Upsilon _{\vartheta}}||^4||{\mathcal{C}_{\vartheta}}||}},\frac{{(1 - 2\varepsilon )\rho }}{{||\Pi_{\vartheta}^ \top {\Upsilon _{\vartheta}}||^2}}} \!\right\}}} \le \frac{\delta }{2} \nonumber
\end{align}
which together with Proposition \ref{ptopk1a} imply that when \eqref{nc1} holds:
\begin{align}
\mathbf{P}( {\mathfrak{E}_2^\mathrm{c}} ) \leq \frac{\delta}{2}. \label{kpp1}
\end{align}

\subsubsection*{\underline{Upper Bound  on $\mathbf{P}( {{\mathfrak{E} _1}\bigcap {{\mathfrak{E} _2}} } )$}} When ${\mathfrak{E}_2}$ occurs, we have
\begin{align}
{\frac{\varrho_{1} - \rho }{\varrho_{2}}}\mathbf{I}_{n} \le {\Psi^\top_{\vartheta}} {X_{\vartheta}}X^\top_{\vartheta} {\Psi_{\vartheta}} \le {\frac{\varrho_{1} + \rho}{\varrho_{2}} }\mathbf{I}_{n}, \nonumber
\end{align}
where $0 < \rho < \varrho_{1}$. We then have
\begin{align}
{\frac{\varrho_{1} - \rho }{\varrho_{2}}}\Psi^{-2}_{\vartheta} \le {X_{\vartheta}}X_{\vartheta}^\top \le {\frac{\varrho_{1} + \rho}{\varrho_{2}} }\Psi^{-2}_{\vartheta}, \label{fg1}
\end{align}
which implies
\begin{align}
\lambda^{0.5}_{\min}({X_{\vartheta}}X_{\vartheta}^\top) \geq \lambda^{0.5}_{\min}({\frac{\varrho_{1} - \rho }{\varrho_{2}}}{\Psi^{-2}_{\vartheta}}) \triangleq \breve{\beta}, \label{fg1j0}
\end{align}
by which we obtain $\frac{1}{\breve{\beta }} \ge || {{{( {{X_{\vartheta}}X_{\vartheta}^ \top })}^{ - 0.5}}} ||$. We then conclude from \eqref{event1} and \eqref{event2} that
\begin{align}
{\mathfrak{E}_1} \bigcap {\mathfrak{E}_2} \subseteq \left\{ {|| {{U}}_{\vartheta}{X}^{\top}_{\vartheta}{({X}_{\vartheta}{X}^{\top}_{\vartheta})^{ - 0.5}} || > \breve{\beta} \phi } \right\} \bigcap {\mathfrak{E}_2}. \label{fg2}
\end{align}

The left-hand inequality of \eqref{fg1} implies
\begin{align}
2{X_{\vartheta}}X_{\vartheta}^ \top  \ge {\frac{\varrho_{1} - \rho }{\varrho_{2}}}\Psi_{\vartheta}^{ - 2} + {X_{\vartheta}}X_{\vartheta}^\top, \nonumber
\end{align}
which means, with $\textcolor[rgb]{0.00,0.00,1.00}{0 < \rho < \rho_{1}}$, that
\begin{align}
{( {{X_{\vartheta}}X_{\vartheta}^ \top })^{ - 1}} \le 2{( {{\frac{\varrho_{1} - \rho }{\varrho_{2}}}\Psi_{\vartheta}^{ - 2} + {X_{\vartheta}}X_{\vartheta}^ \top })^{ - 1}}, \nonumber
\end{align}
which, in conjunction with \eqref{fg2}, leads to
\begin{align}
{\mathfrak{E}_1} \!\bigcap\! {{\mathfrak{E}_2}} \subseteq\!\! \left\{\!\! {\sqrt 2 || {{{U}}_{\vartheta}{X}^{\top}_{\vartheta}{{( {{S} \!+\! {X_{\vartheta}}X_{\vartheta}^ \top })}^{ - 0.5}}} || \!>\! \breve{\beta} \phi } \!\right\} \!\bigcap\! {{\mathfrak{E}_2}},\label{fg3}
\end{align}
where we denote
\begin{align}
S \triangleq {\frac{\varrho_{1} - \rho }{\varrho_{2}}}\Psi_{\vartheta}^{ - 2}. \label{kks}
\end{align}

With $\mathbf{u} \in \mathcal{S}^{n-1}$, we now define two additional events:
\begin{align}
&{\mathfrak{A}_1} \triangleq \left\{{{{|| {{{( {S + {X_{\vartheta}}X_{\vartheta}^\top })^{ - 0.5}}}X_{\vartheta}}{{U}}^{\top}_{\vartheta} ||}^2}} \right. \nonumber\\
&\hspace{1.00cm}\left. { > 16\mathfrak{c}{\kappa ^2}\ln ( {{{( {\det( {( {S + {X_{\vartheta}}X_{\vartheta}^ \top }){S^{ - 1}}})} )}^{0.5}}\delta^{-1}_0})} \right\}, \label{adf1}\\
&{\mathfrak{A}_2}(\mathbf{u}) \triangleq \left\{ {{{|| {{{( {S + {X_{\vartheta}}X_{\vartheta}^\top })}^{ - 0.5}}X_{\vartheta}{{U}}^{\top}_{\vartheta} \mathbf{u}} ||_{2}^2}}} \right.\nonumber\\
&\hspace{1.00cm}\left. { > 4\mathfrak{c}{\kappa ^2}\ln({{{({\det( {( {S + {X_{\vartheta}}X_{\vartheta}^\top }){S^{ - 1}}})})}^{0.5}}\delta^{-1}_0})} \right\}. \label{adf2}
\end{align}

We note that under Assumption \ref{asprv}-3), $\mathbf{u} \in \mathcal{S}^{n-1}$ implies that $(\widehat{\mathfrak{g}}^{j+1}_{k+1})^{\top}\mathbf{u}$ is $\mathcal{F}_{k}$-measurable and conditionally $\gamma$-sub-Gaussian for some $\gamma > 0$. Meanwhile, we note that $4\mathfrak{c}{\kappa^2} = 2{\gamma^2}$. In light of  Lemma \ref{ckpq} in Appendix B,  we then have $\mathbf{P}[{\mathfrak{A}_2}(\mathbf{u})] \leq \delta_{0}$. Furthermore, applying \eqref{aalem1} with the setting of $\widehat{\varepsilon} = \frac{{2 + \overline{\varepsilon} }}{\overline{\varepsilon} }$ with $\overline{\varepsilon} \in [0,1)$  in Lemma \ref{cko}, we obtain
\begin{align}
\mathbf{P}[{\mathfrak{A}_1}] \leq {\widehat{\varepsilon}^n}\mathop {\max }\limits_{\mathbf{u} \in \mathcal{N}} \mathbf{P}[ {{\mathfrak{A}_2(\mathbf{u})}}] \le {\widehat{\varepsilon}^n}{\delta _0}.\label{fg4}
\end{align}

We let  ${\delta _0} = \frac{\delta }{{2 \cdot {\widehat{\varepsilon}^n}}}$, such that
\begin{align}
\breve{\beta}  &\ge \frac{{4\sqrt {2\mathfrak{c}} \kappa }}{\phi }\sqrt {\ln \left(\frac{{2 \cdot \widehat{\varepsilon}^n}}{\delta}{{\left( {\frac{{2\varrho_{1}}}{{\varrho_{1} - \rho}}} \right)}^{0.5n}}\right)} \nonumber\\
 &\!=\! \frac{{4\sqrt {2\mathfrak{c}} \kappa }}{\phi }\sqrt {\ln \left(\frac{{2 \cdot {\widehat{\varepsilon}^n}}}{{{2\cdot\delta _0} \cdot {\widehat{\varepsilon}^n}}}{{\left( {\frac{2\varrho_{1}}{{\varrho_{1} - \rho }}} \right)}^{0.5n}}\right)} \nonumber\\
 &\!=\! \frac{{4\sqrt {2\mathfrak{c}} \kappa }}{\phi }\sqrt {\ln \left( {\frac{1}{{{\delta _0}}}{{\left( {\det \left( {\left( {S + \textcolor[rgb]{0.00,0.00,1.00}{\frac{{\varrho_{1} +\rho }}{{\varrho_{1} - \rho}}S}} \right){S^{ - 1}}} \right)} \right)}^{0.5}}} \right)} \nonumber\\
 &\!\geq\! \frac{{4\sqrt {2\mathfrak{c}} \kappa }}{\phi }\sqrt {\ln \left( {\frac{1}{{{\delta _0}}}\!{{\left( {\det({( {S + {X_{\vartheta}}X_{\vartheta}^ \top }){S^{ - 1}}})} \right)^{0.5}}}} \right)},\label{pq}
\end{align}
where the last inequality from its previous step is obtained via considering the inequality $\textcolor[rgb]{0.00,0.00,1.00}{{X_{\vartheta}}X_{\vartheta}^ \top  \leq \frac{{\varrho_{1} + \rho }}{{\varrho_{1} - \rho }}S}$ that follows from \eqref{kks} and the right-hand inequality of \eqref{fg1}.

Combining the inequality in \eqref{fg3} with \eqref{pq} yields
\begin{align}
&|| {{{U}}_{\vartheta}{X}^{\top}_{\vartheta}{{( {S + {X_{\vartheta}}X_{\vartheta}^ \top })}^{ - 0.5}}} || \nonumber\\
& > \frac{{\breve{\beta }\phi }}{{\sqrt 2 }} \ge 4\sqrt \mathfrak{c} \kappa \sqrt {\ln \left( {\frac{1}{{{\delta _0}}}{{( {\det(( {S + {X_{\vartheta}}X_{\vartheta}^ \top ){S^{ - 1}}})})^{0.5}}}}\right)}, \nonumber
\end{align}
by which, and considering \eqref{fg3} and \eqref{adf1}, we deduce that \underline{under condition \eqref{pq}}, if the event ${\mathfrak{E}_1}$ given in \eqref{event1} occurs, the event ${\mathfrak{A}_1}$ given in \eqref{adf1} occurs consequently. We thus obtain
\begin{align}
\mathbf{P}[{\mathfrak{E}_1}\bigcap {{\mathfrak{E}_2}} ]  \leq \mathbf{P}[{{\mathfrak{A}_1}}\bigcap {{\mathfrak{E}_2}}].\label{fg4pk}
\end{align}

With the consideration of $\widehat{\varepsilon} = \frac{{2 + \overline{\varepsilon} }}{\overline{\varepsilon} }$, we conclude that the condition \eqref{nc2} is equivalent to
\begin{align}
\lambda _{\min }^{0.5}\!\left(\! {\frac{\varrho_{1} \!-\! \rho }{\varrho_{2}}{\Gamma _{\vartheta}}} \!\right) \ge \frac{{4\sqrt {2\mathfrak{c}} \kappa }}{\phi }\sqrt {\ln\! \left(\frac{{2 \cdot \widehat{\varepsilon}^n}}{\delta}{{\left( {\frac{{2\varrho_{1}}}{{\varrho_{1} \!-\! \rho}}} \right)}^{0.5n}}\right)} > 0, \nonumber
\end{align}
inserting $\breve{\beta}$ in \eqref{fg1j0} with $\Gamma _{( {k,p})}$ in \eqref{oEp2} into which yields
\begin{align}
\breve{\beta}  \ge \frac{{4\sqrt {2\mathfrak{c}} \kappa }}{\phi }\sqrt {\ln \!\left(\frac{{2 \cdot \widehat{\varepsilon}^n}}{\delta}{{\left( {\frac{{2\varrho_{1}}}{{\varrho_{1} - \rho}}} \right)}^{0.5n}}\right)} > 0, \nonumber
\end{align}
by which we conclude that \eqref{pq} holds if the condition \eqref{nc2} is satisfied. Moreover, recalling that the event $\mathfrak{E} _2$ always occurs under the condition \eqref{nc1} (proved in \underline{Upper Bound on $\mathbf{P}( {\mathfrak{E} _2^\mathrm{c}})$}), we conclude from \eqref{fg4pk} and \eqref{fg4} that
\vspace{-0.1cm}
\begin{align}
\mathbf{P}[ {{\mathfrak{E} _1}\bigcap {{\mathfrak{E} _2}} }]  \leq \mathbf{P}[ {\mathfrak{A}_1}] \leq {\widehat{\varepsilon}^n}{\delta _0} \nonumber
\end{align}
holds as long as both \eqref{nc1} and \eqref{nc2} hold. In addition, due to ${\delta _0} = \frac{\delta }{{2 \cdot {\widehat{\varepsilon}^n}}}$, we have
\vspace{-0.3cm}
\begin{align}
\mathbf{P}[ {{\mathfrak{E}_1}\bigcap {{\mathfrak{E}_2}} }]  \leq  \frac{\delta}{2}. \label{fg4pk2}
\end{align}
Finally, combining \eqref{kp1} with \eqref{kpp1} and \eqref{fg4pk2} yields \eqref{fnfn}.

\section*{Appendix E: Proof of Corollary \ref{fth}}
\subsubsection*{\underline{Condition \eqref{fcc1}}} \textcolor[rgb]{0.00,0.00,1.00}{Let us denote:
\begin{align}
\widetilde{\mathbf{z}}^{j}_{g} &\!\triangleq\! ( {\mathcal{M}_{g}^\mathrm{c} \!-\! \mathcal{M}_{j}^\mathrm{c}})x(1)  \!+\! \sum\limits_{i = 1}^{g - 1} {\mathcal{M}_{(i,g)}}(\mathbf{I}_{n} \!-\! \mathcal{M}_{(j - g+1,j)}) {{\alpha _{\vartheta(g-i)}}}  \nonumber\\
&\hspace{4.4cm}   - \sum\limits_{i = 1}^{j - g}\!{\mathcal{M}_{(i,j)}}{{\alpha _{\vartheta(j-i)}}},\label{opop1}\\
\widetilde{\mathbf{p}}^{j}_{g} &\!\triangleq\! \widetilde{\mathfrak{o}}_g^j  + \sum\limits_{i = 1}^{g - 1} {\mathcal{M}_{(i,g)}}(\mathbf{I}_{n} \!-\! \mathcal{M}_{(j - g+1,j)})\mathfrak{p}(g-i)\nonumber\\
&\hspace{4.3cm} - \sum\limits_{i = 1}^{j - g}{\mathcal{M}_{(i,j)}}\mathfrak{p}(j\!-\!i),\label{opop2}
\end{align}
such that \eqref{tr1} can be rewritten as $\widetilde{\mathbf{y}}^{j}_{g} = \widetilde{\mathbf{z}}^{j}_{g} + \widetilde{\mathbf{p}}^{j}_{g}$. We observe from \eqref{opop1} and \eqref{opop2} that $\widetilde{\mathbf{z}}^{j}_{g}$ is a deterministic vector while $\widetilde{\mathbf{p}}^{j}_{g}$ is a random vector. Under Assumption \ref{asprv}-1), we then obtain from \eqref{opop1}, \eqref{opop2}, \eqref{traaab} and \eqref{nc1dab} that}
\begin{align}
&\mathbf{E}[\widetilde{\mathbf{y}}^{j}_{g}(\widetilde{\mathbf{y}}^{j}_{g})^{\top}] \nonumber\\
&\textcolor[rgb]{0.00,0.00,1.00}{\!=\! (\widetilde{\mathbf{z}}^{j}_{g})^\top\widetilde{\mathbf{z}}^{j}_{g} + \mathbf{E}[(\widetilde{\mathbf{p}}^{j}_{g})^{\top}\widetilde{\mathbf{p}}^{j}_{g}]} \nonumber\\
&\textcolor[rgb]{0.00,0.00,1.00}{\!\geq\! \mathbf{E}[(\widetilde{\mathbf{p}}^{j}_{g})^{\top}\widetilde{\mathbf{p}}^{j}_{g}]} \nonumber\\
&\!=\! \mathbf{E}[\widetilde{\mathfrak{o}}_g^j(\widetilde{\mathfrak{o}}_g^j)\!^{\top}] \!+\!\! \sum\limits_{i = 1}^{g-1}\!{\mathcal{M}_{\!(i,g)}}\!(\mathbf{I}_{n} \!-\!\! \mathcal{M}_{\!(j - g+1,j)}\!)\mathbf{E}[\mathfrak{p}(g\!-\!j)\mathfrak{p}\!^{\top}\!\!(g\!-\!j)] \nonumber\\
&\cdot\!(\mathbf{I}_{n} \!-\! \mathcal{M}_{\!(j - g+1,j)}\!)\!^{\top}\!\!{\mathcal{M}^{\top}_{\!(i,g)}} \!+\!\!  \sum\limits_{i = 1}^{j - g}\!{\mathcal{M}_{(i,j)}}\!\mathbf{E}[\mathfrak{p}(j\!-\!i)\mathfrak{p}\!^{\top}\!\!(j\!-\!i)]\mathcal{M}^{\top}_{\!(i,j)}  \nonumber\\
&\geq 2\sigma^{2}_{\mathrm{o}}\mathbf{I}_{n} + (j - 1)\sigma^{2}_{\mathrm{p}}{\underline{\mathfrak{s}}} \mathbf{I}_{n}, \nonumber
\end{align}
which together with \eqref{oEp2} lead to ${\Gamma _{\vartheta}} \ge 2\frac{{\mathfrak{l}_{\vartheta}}}{n}\sigma^{2}_{\mathrm{o}}\mathbf{I}_{n} + \underline{\mathfrak{s}}\frac{{\overline{\mathfrak{l}}_{\vartheta}}}{n}\sigma^{2}_{\mathrm{p}} \mathbf{I}_{n}$ where $\mathfrak{l}_{\vartheta}$ and $\overline{\mathfrak{l}}_{\vartheta}$ are given in \eqref{ghcc} and \eqref{ghccan}, respectively. As a consequence, we have
\begin{align}
{\lambda_{\min }}( \Gamma _{\vartheta}) \ge 2\frac{{\mathfrak{l}_{\vartheta}}}{n}\sigma^{2}_{\mathrm{o}} + \underline{\mathfrak{s}}\frac{{\overline{\mathfrak{l}}_{\vartheta}}}{n}\sigma^{2}_{\mathrm{p}} = \mathfrak{f}_{\vartheta},\label{tkc}
\end{align}
which implies that \eqref{nc2} holds if \eqref{fcc1} is satisfied.

\subsubsection*{\underline{Condition \eqref{fccq}}} Considering \eqref{nc1dcc}, it follows from \eqref{adef1} and \eqref{gh9} with \eqref{gh2} and \eqref{ghmm2} that
${|| {{\Upsilon}_{\vartheta}} ||^2}  = {|| \Psi _{\vartheta} ||^2}$. We obtain from \eqref{oEp2} and \eqref{tkc} that $\Psi _{\vartheta}^2 \le \frac{{{{\bf{I}}_n}}}{2\frac{{\mathfrak{l}_{\vartheta}}}{n}\sigma^{2}_{\mathrm{o}} + \underline{\mathfrak{s}}\frac{{\overline{\mathfrak{l}}_{\vartheta}}}{n}\sigma^{2}_{\mathrm{p}}}$, which thus leads to
\begin{align}
 {|| {{\Upsilon}_{\vartheta}} ||^2}  \le \frac{1}{2\frac{{\mathfrak{l}_{\vartheta}}}{n}\sigma^{2}_{\mathrm{o}} + \underline{\mathfrak{s}}\frac{{\overline{\mathfrak{l}}_{\vartheta}}}{n}\sigma^{2}_{\mathrm{p}}}. \label{pkq}
\end{align}

Considering \eqref{nc1dcc} and \eqref{pkq} and recalling the well-known inequality $\left\| AB \right\| \le \left\| A \right\|\left\| B \right\|$, we arrive at
\begin{align}
||{{\Pi}^{\top}_{\vartheta}}{{\Upsilon}_{\vartheta}} ||^{2} &\leq ||{{\Pi}_{\vartheta}}||^2||{{\Upsilon}_{\vartheta}}||^2  \leq  {\mathfrak{j}_{\vartheta}},\label{ckmq}
\end{align}
where ${\mathfrak{j}_{\vartheta}}$ is given in \eqref{nad1tt}. We note that \eqref{ckmq} implies
\begin{align}
&\min\! \left\{\! {\frac{{{{(1 - 2\varepsilon )}^2}n{\rho ^2}}}{{\mathfrak{l}_{\vartheta}||\Pi _{\vartheta}^\top {\Upsilon _{\vartheta}}||^{4}||{\mathcal{C}_{\vartheta}}||}},~\frac{{(1 - 2\varepsilon )\rho }}{{||\Pi_{\vartheta}^ \top {\Upsilon _{\vartheta}}||^2}}} \!\right\} \nonumber\\
&\geq  \min\! \left\{\! {\frac{{{{(1 - 2\varepsilon)}^2}n{\rho ^2}}}{{\mathfrak{l}_{\vartheta}{\mathfrak{j}^{2}_{\vartheta}}||{\mathcal{C}_{\vartheta}}||}},~\frac{{(1 - 2\varepsilon )\rho }}{{{\mathfrak{j}_{\vartheta}}}}} \!\right\},\nonumber
\end{align}
which indicates that if \eqref{fccq} is satisfied, the \eqref{nc1} holds.

\section*{Appendix F: Proof of Theorem \ref{th2}}
We obtain from \eqref{dec2} that
\begin{align}
&[\widehat{\mathfrak{a}}_{+1}]_{i} - [\widehat{\mathfrak{a}}_{-1}]_{i} = 2( { \breve{\epsilon}_{i} + {\breve{\eta}_i}}), \label{apr1}\\
&[\widehat{\mathfrak{a}}_{+1}]_{i} + [\widehat{\mathfrak{a}}_{-1}]_{i} = 2( {1 - \sum\limits_{j \in \mathbb{V}} {{\breve{w}_{ij}}}}){\breve{s}_i} - 2( {{\breve{\epsilon}_i} + {\breve{\eta}_i}}){\breve{s}_i}, \label{apr2}\\
&\sum\limits_{j \in \mathbb{V}} {( {{{[ \widehat{\mathcal{A}}_{+1}]}_{i,j}} - {{[ \widehat{\mathcal{A}}_{-1}]}_{i,j}}} )}  = 2({\breve{\eta}_i} - {\breve{\epsilon}_i}){\breve{s}_i},\label{apr3}\\
&\sum\limits_{j \in \mathbb{V}} {( {{{[ \widehat{\mathcal{A}}_{+1}]}_{i,j}} + {{[ \widehat{\mathcal{A}}_{-1}]}_{i,j}}} )}  =  2\sum\limits_{j \in \mathbb{V}} {{\breve{w}_{ij}}} - 2{\breve{\eta}_i} + 2{\breve{\epsilon}_i}.\label{apr4}
\end{align}
Combining \eqref{apr1} with \eqref{apr2} yields
\begin{align}
\frac{1}{\breve{s}_i}([\widehat{\mathfrak{a}}_{+1}]_{i} \!+\! [\widehat{\mathfrak{a}}_{-1}]_{i}) \!=\! 2( {1 \!-\! \sum\limits_{j \in \mathbb{V}} {{\breve{w}_{ij}}}}) \!-\! ([\widehat{\mathfrak{a}}_{+1}]_{i} \!-\! [\widehat{\mathfrak{a}}_{-1}]_{i} ). \label{apr5}
\end{align}
Meanwhile, combining \eqref{apr3} with \eqref{apr4} leads to
\begin{align}
\frac{1}{\breve{s}_i}\!\sum\limits_{\!j \in \mathbb{V}} \!\!{( {{{[ \widehat{\mathcal{A}}_{+1}]}_{i,j}} \!-\! {{[ \widehat{\mathcal{A}}_{-1}]}_{i,j}}} \!)} \!=\! 2\!\sum\limits_{\!j \in \mathbb{V}} \!{{\breve{w}_{ij}}} \!-\!\! \sum\limits_{\!j \in \mathbb{V}}\!\! {( {{{[ \widehat{\mathcal{A}}_{+1}]}_{i,j}} \!+\! {{[ \widehat{\mathcal{A}}_{-1}]}_{i,j}}} \!)}, \nonumber
\end{align}
adding \eqref{apr5} into which, we arrive at
\begin{align}
&\frac{1}{\breve{s}_i}([\widehat{\mathfrak{a}}_{+1}]_{i} + [\widehat{\mathfrak{a}}_{-1}]_{i} + \sum\limits_{j \in \mathbb{V}} \!{( {{{[ \widehat{\mathcal{A}}_{+1}]}_{i,j}} \!-\! {{[ \widehat{\mathcal{A}}_{-1}]}_{i,j}}} )}) \nonumber\\
&\!=\! 2 - ([\widehat{\mathfrak{a}}_{+1}]_{i} - [\widehat{\mathfrak{a}}_{-1}]_{i} ) -  \sum\limits_{j \in \mathbb{V}}\! {( {{{[ \widehat{\mathcal{A}}_{+1}]}_{i,j}} + {{[ \widehat{\mathcal{A}}_{-1}]}_{i,j}}} )}, \nonumber
\end{align}
which results in the computation of $\breve{s}_{i}$ in Line 1 of Algorithm 1. With computed $\breve{s}_{i}$, from \eqref{apr1} and \eqref{apr3} we have
\begin{align}
[\widehat{\mathfrak{a}}_{+1}]_{i} - [\widehat{\mathfrak{a}}_{-1}]_{i} - \frac{1}{\breve{s}_{i}}\sum\limits_{j \in \mathbb{V}} {( {{{[ \widehat{\mathcal{A}}_{+1}]}_{i,j}} - {{[ \widehat{\mathcal{A}}_{-1}]}_{i,j}}} )} = 4\breve{\epsilon}_{i}, \nonumber\\
[\widehat{\mathfrak{a}}_{+1}]_{i} - [\widehat{\mathfrak{a}}_{-1}]_{i} + \frac{1}{\breve{s}_{i}}\sum\limits_{j \in \mathbb{V}} {( {{{[ \widehat{\mathcal{A}}_{+1}]}_{i,j}} - {{[ \widehat{\mathcal{A}}_{-1}]}_{i,j}}} )} = 4\breve{\eta}_{i}, \nonumber
\end{align}
which indicates the computations of $\breve{\epsilon}_{i}$ and $\breve{\eta}_{i}$ in in Lines 2 and 3 of Algorithm 1, respectively. With the obtained $\breve{\epsilon}_{i}$ and $\breve{\eta}_{i}$, the relation \eqref{apr4}  implies the computation of the sum of social-influence weights in Line 4 of Algorithm 1. We obtain from \eqref{dec2a} and \eqref{dec2b}  that
\begin{align}
{[\widehat{\mathcal{A}}_{+1}]_{i,j}} + {[\widehat{\mathcal{A}}_{-1}]_{i,j}} = \begin{cases}
\!2{(1 - \frac{{{\breve{\eta}_i}}}{{\sum\limits_{l \in \mathbb{V}} {{{{{\breve{w}_{il}}}}}} }})\breve{w}_{ii} + 2{\breve{\epsilon}_i}} &\text{if}~i \!=\! j \\
\!  2(1 - \frac{{{\breve{\eta}_i}}}{{\sum\limits_{l \in \mathbb{V}} {{{{{\breve{w}_{il}}}}}} }}){\breve{w}_{ij}},&\text{if}~i \!\neq\! j
\end{cases} \nonumber
\end{align}
which, with the computed ${\breve{\epsilon}_i}$ and the sum $\sum\limits_{j \in \mathbb{V}} {{\breve{w}_{ij}}}$, lead to the computation of weighted network topology in Line 5 of Algorithm 1.

\vspace{-0.0cm}
\bibliographystyle{IEEEtran}
\bibliography{ref}
\end{document}